\documentclass[manuscript]{acmart}

\AtBeginDocument{%
  }

\acmConference[Conference acronym 'XX]{}{June 03--05,
  2024}{Woodstock, NY}

\usepackage{csquotes}
\usepackage{graphicx}
\usepackage{subcaption}
\usepackage{wrapfig}
\usepackage{tabularx}
\begin{document}

\title{Farmer.Chat: Scaling AI-Powered Agricultural Services for Smallholder Farmers}
\author{Namita Singh, Jacqueline Wang'ombe, Nereah Okanga, Tetyana Zelenska, Jona Repishti, Jayasankar G K, Sanjeev Mishra, Rajsekar Manokaran, Vineet Singh, Mohammed Irfan Rafiq, Rikin Gandhi}
\affiliation{%
  \institution{Digital Green}
  \country{India}}

\author{Akshay Nambi}
\affiliation{%
  \institution{Microsoft Research}
  \country{India}
 }
\email{akshayn@microsoft.com}


\renewcommand{\shortauthors}{Singh et al.}

\begin{abstract}
Small and medium-sized agricultural holders face challenges like limited access to localized, timely information, impacting productivity and sustainability. Traditional extension services, which rely on in-person agents, struggle with scalability and timely delivery, especially in remote areas. We introduce ``Farmer.Chat," a generative AI-powered chatbot designed to address these issues. Leveraging Generative AI, Farmer.Chat offers personalized, reliable, and contextually relevant advice, overcoming limitations of previous chatbots in deterministic dialogue flows, language support, and unstructured data processing. Deployed in four countries, Farmer.Chat has engaged over 15,000 farmers and answered over 300,000 queries. This paper highlights how Farmer.Chat’s innovative use of GenAI enhances agricultural service scalability and effectiveness. Our evaluation, combining quantitative analysis and qualitative insights, highlights Farmer.Chat’s effectiveness in improving farming practices, enhancing trust, response quality, and user engagement. 
\end{abstract}

\maketitle

\section{Introduction}
Smallholder farmers are essential to global food security but often struggle to access timely, localized, and actionable agricultural information~\cite{faure2012new,fabregas2019realizing,zerssa2021challenges}. This gap hinders their productivity, profitability, and adoption of sustainable practices, particularly in low- and middle-income countries where farming technologies require local adaptation based on plot-specific factors~\cite{davis2014role,fan2020role}. The lack of reliable guidance limits farmers’ ability to assess risks and leads to lower yields and incomes~\cite{anderson2008agricultural,myeni2019barriers,jama2008agriculture}. 

Agricultural extension services aim to bridge this gap by providing farmers with knowledge and resources~\cite{anderson2007agricultural,swanson2005improving,saliu2009trends}. However, traditional services depend on extension agents who face challenges in delivering timely, tailored advice, especially in remote areas where factors like soil, climate, and crop variety differ greatly, making it hard to provide precise recommendations~\cite{ferroni2012achievements,raabe2008reforming,nedumaran2019agriculture,msuya2017role}. 

Recent efforts have explored information and communication technologies (ICTs)~\cite{spielman2021information, tata2018impact, saravanan2010icts,sanga2013building,islam2017utilization}, such as SMS reminders~\cite{silvestri2021going,kachelriess2018using,larochelle2019did} and video learning programs~\cite{gandhi2007digital,david2011video}, to enhance agricultural extension. While promising, these methods rely on curated content and human mediation, limiting scalability and adaptability. There is a growing need for conversational systems that can process diverse data and handle various inquiries, particularly for low-literacy users who require natural language interactions and access to dynamic knowledge bases~\cite{saravanan2010icts}. Traditional chatbots and models often fall short due to high resource demands and the need for human intervention~\cite{saravanan2010icts,gandhi2007digital}.

Large Language Models (LLMs) powered by Generative AI (GenAI) offer a transformative solution~\cite{silva2023gpt,DGOpenAI}. Using Retrieval-Augmented Generation (RAG)~\cite{fan2024survey}, LLMs can process unstructured data—such as documents, crop tables, and videos—providing contextually relevant, accurate, and personalized information. These systems create a seamless user experience through natural dialogue, guiding farmers with tailored, adaptive interactions while ensuring the advice is trustworthy and actionable.

We present Farmer.Chat, a generative AI-powered chatbot designed to address key challenges in traditional agricultural extension services. Offering a scalable and customizable solution, it provides smallholder farmers with timely, reliable, and context-specific information. Unlike prior chatbots~\cite{venkata2022farmer,momaya2021krushi,jain2018farmchat} that rely on human mediation, Farmer.Chat delivers on-demand, tailored agricultural advice across multiple platforms, including messaging apps and mobile applications. This design overcomes geographical limitations, scalability issues, and dependence on human intermediaries, enhancing access to agricultural knowledge for marginalized groups like women and low-literacy users.

Farmer.Chat leverages RAG technology to process structured and unstructured data—such as research papers, crop tables, and videos—building a dynamic and adaptable knowledge base. With multilingual support and multimedia responses (text, audio, video), it ensures accessibility for diverse users, particularly those with low literacy. Localized knowledge bases ensure region-specific, accurate advice tailored to local agricultural practices. 

Deployed in four countries—Kenya, India, Ethiopia, and Nigeria—Farmer.Chat engages over 15,000 farmers and has addressed more than 300,000 queries. Available in six languages, it serves diverse communities, covering over 40 value chain crops, and demonstrates its capacity to scale globally while providing relevant, accessible agricultural support.

This paper focuses on our Kenya deployment, started in October 2023, which now has the largest user base of 8,805 users across seven counties. These users are categorized into four groups:
\begin{enumerate}
    \item \textbf{Agriculture Extension Agents (EAs):} Employed by Kenya's Ministry of Agriculture, EAs disseminate agricultural knowledge but face high farmer-to-agent ratios—ranging from 1:1,000 to 1:5,000~\cite{ilukor2019investments,odhong2018private}—making it challenging to reach all farmers.
    \item \textbf{Lead Farmers:} To bridge this gap, a farmer-to-farmer extension model selects Lead Farmers based on experience and leadership, each supporting 15-30 farmers with information from EAs.
    \item \textbf{Farmers:} The main beneficiaries, receiving personalized agricultural advice via Farmer.Chat to enhance their practices.
    \item \textbf{Agripreneurs:} A developing group introducing innovative technologies and market access, driving agricultural transformation.
\end{enumerate}
This paper addresses the following core research questions: \textbf{(RQ1.)} How can generative AI enhance the scalability, accessibility, and contextual relevance of agricultural extension services? \textbf{(RQ2.)} What are the key factors influencing user trust, engagement, and long-term adoption of AI-driven agricultural advisory tools like Farmer.Chat? \textbf{(RQ3.)} How does the deployment of Farmer.Chat impact agricultural outcomes, farmer satisfaction, and community-level adoption across diverse regions?

The key contributions of this paper are threefold. First, we introduce Farmer.Chat, a novel AI-driven agricultural extension platform that uses Generative AI to deliver multilingual, multimodal interactions and contextually relevant, scalable agricultural advice. The platform seamlessly integrates structured and unstructured data, providing timely, personalized recommendations tailored to diverse user needs. Second, through qualitative user studies involving 300+ users across focus groups, interviews, usability tests, and surveys, we examine user experience, trust, and usability. Results show high practicality, ease of interaction, and enhanced user agency, particularly for women and low-literacy farmers, emphasizing Farmer.Chat's ability to build trust and deliver actionable information. Third, quantitative analysis reveals strong system performance, with over 75\% of queries successfully answered. Users report high satisfaction with response quality and relevance, showing strong engagement on critical topics like yield, pest control, and weather.

\section{Related Work}
In this section, we review research on (1) Agricultural Extension Services and ICT Interventions, (2) Chatbots and Conversation Agents and (3) Generative AI in Agriculture, thus contextualizing the key novelty of Farmer.Chat.  

\subsection{Agricultural Extension Services and ICT Interventions}
Traditional agricultural extension services have been essential in disseminating knowledge to farmers~\cite{anderson2007agricultural,saliu2009trends}. However, in many low- and middle-income countries, they face significant challenges, such as limited reach due to an insufficient number of extension agents~\cite{ferroni2012achievements,msuya2017role}. In Kenya, for example, the agent-to-farmer ratio is estimated at 1:1000 according to government reports (NASEP 2012~\cite{NASEP}), but it can be as high as 1:4,000~\cite{franzel2014farmer,muyanga2006agricultural}, far below the recommended ratio of 1:400. Additionally, these services tend to be top-down, limiting farmer input and reducing engagement~\cite{msuya2017role}. While peer-learning models like farmer field schools aim to address these gaps, they often face resource limitations and inconsistent results due to varying farming contexts such as soil, climate, and crop variety~\cite{waddington2014farmer,feder2004sending,waddington2014farmer1}.

In response to these limitations, ICT-based interventions have been developed to scale agricultural knowledge dissemination~\cite{saravanan2010icts,spielman2021information}. Mobile solutions like SMS and voice-based systems have shown promise in delivering timely, localized information. In Kenya, SMS reminders increased sugarcane yields by 8\% and boosted revenue by US\$54 per farmer at a cost of just US\$0.02 per message~\cite{van2021information}. Similarly, a voice-based system for cotton farmers in India demonstrated high adoption rates, especially among low-literacy users~\cite{patel2012power,patel2010avaaj}. Despite their success, these ICT solutions rely on curated content, limiting adaptability to changing agricultural conditions~\cite{saravanan2014mobile}.

Peer-to-peer video learning models have also improved adoption rates by using locally produced, farmer-led videos to share context-specific knowledge~\cite{gandhi2007digital}. Studies show video-enabled extension is ten times more cost-effective than traditional services, reducing the cost per adoption from US\$35 to US\$3.50 and reaching 30\% more farmers~\cite{DG}. However, producing high-quality, localized videos is resource-intensive, and reliance on human facilitators limits scalability~\cite{ibeawuchi2021review}.

While ICT and peer-learning interventions have enhanced agricultural extension, scalability and adaptability challenges persist. AI-driven solutions like Farmer.Chat can address these issues by providing scalable, cost-effective, and context-aware support, meeting the growing need for effective agricultural extension services.

\subsection{Chatbots and Conversation Agents in Agriculture}
Chatbots and conversational agents are increasingly used in agriculture for providing accessible information through natural language interactions. Projects like Hello Tractor and Avaaj Otalo offer real-time advice on topics such as weather, pest control, and farming techniques via voice or text systems~\cite{patel2010avaaj, patel2012power,jain2018farmchat}.

While rule-based chatbots are useful for structured, repetitive tasks, they struggle with complex, dynamic queries that require context-awareness. Systems like Avaaj Otalo handle voice queries but are limited in adapting to evolving agricultural needs~\cite{patel2010avaaj}. Similarly, FarmChat faces challenges in offering personalized advice due to variables like soil type, climate, and crop variety~\cite{jain2018farmchat}, and most traditional chatbots lack real-time updates and access to diverse data sources, reducing their effectiveness in dynamic agricultural environments~\cite{darapaneni2022farmer}.

In contrast, AI-driven chatbots use machine learning and natural language processing (NLP) to deliver more flexible, personalized, and data-driven responses. Studies show that AI systems outperform rule-based chatbots in user satisfaction, contextual understanding, and scalability~\cite{adamopoulou2020overview, shawar2007chatbots, serban2015survey}. AI-driven chatbots also adapt to real-time data, providing dynamic, personalized advice~\cite{folstad2017chatbots, klopfenstein2017rise}. Recent work highlights AI's ability to integrate diverse data sources to address complex agricultural needs~\cite{madotto2020language, yildirim2023few, xu2024sa}.

Farmer.Chat builds on these advancements by leveraging AI models to offer personalized, real-time recommendations based on dynamic, context-specific data. Unlike traditional chatbots, Farmer.Chat adapts to changing agricultural conditions and provides tailored advice.

\subsection{Generative AI in Agriculture}
Advancements in generative AI, particularly large language models (LLMs) like GPT-3 and GPT-4~\cite{achiam2023gpt}, are transforming agricultural knowledge accessibility, especially in low-resource settings. Projects like Kisan.AI~\cite{kissan} have deployed LLMs to offer real-time advice on crop management and pest control~\cite{kissan}. However, these systems face challenges in adapting to diverse agricultural ecosystems due to limited knowledge bases and difficulties in ingesting non-digital agricultural information. Additionally, the lack of robust multilingual support and the inability to handle multimodal inputs, such as images and audio, further restricts their usefulness in rural contexts~\cite{raiaan2024review}.

Most existing LLM-based agricultural chatbots focus on a narrow range of crops and regions, neglecting the complex needs of smallholder farmers~\cite{tzachor2023large,silva2023gpt}. Their inability to integrate localized weather and soil data reduces their precision in providing actionable insights. Farmer.Chat addresses these limitations by supporting multiple crops, integrating real-time weather and soil data, and delivering personalized recommendations. Its multilingual and multimodal capabilities (audio, image, video) make it accessible to low-literacy users, crucial in rural settings. Using Retrieval-Augmented Generation (RAG) for structured and unstructured data, Farmer.Chat enhances trustworthiness and precision.

Finally, designing AI-driven tools for low-literacy, resource-constrained populations requires intuitive, culturally sensitive interfaces. Prior studies~\cite{medhi2006text,patel2010avaaj,parikh2006mobile,sherwani2007healthline} demonstrate the effectiveness of voice-based systems and image-based interfaces for increasing engagement. Further, several research highlights the importance of culturally relevant, trust-building designs for sustainable use~\cite{brewer2005case,bidwell2016moving,jackson2011things}. Jackson et al.\cite{jackson2011things} and Sambasivan et al.\cite{ejiaku2014technology} stress the need for AI tools to align with local practices and function well in resource-limited environments. Dell et al.\cite{dell2012yours} underscore offline functionality, while Amershi et al.\cite{amershi2019guidelines} advocate for clear feedback and user control. These insights shape Farmer.Chat's design, ensuring personalized, accessible, and context-aware support for low-literacy farmers.

\begin{figure}[t!]\centering
    \includegraphics[width=0.95\linewidth]{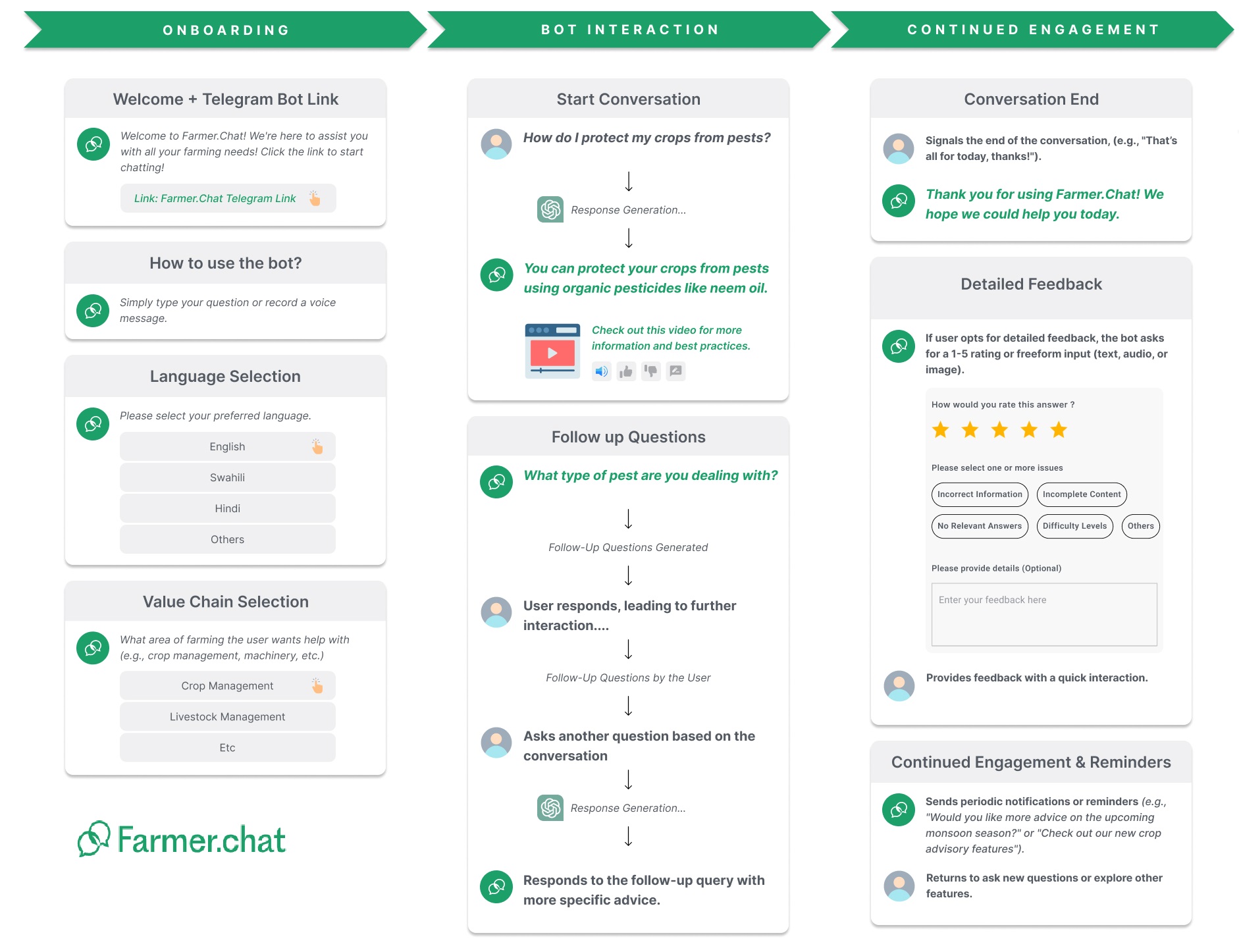}
    \vspace{-10pt}
    \caption{Overview of Interaction Flow in Farmer.Chat.}
    \label{fig:interaction}
    \vspace{-15pt}
\end{figure}
\section{Farmer.Chat Overall Design and Architecture}
\subsection{Overall Design}
The design of Farmer.Chat is grounded in user-centered principles~\cite{amershi2019guidelines,medhi2006text}, tailored for low-literacy populations in rural, resource-constrained environments. It is built on three core pillars: \textbf{ease of use}, \textbf{multilingual and multimodal interactions}, and a \textbf{comprehensive agricultural knowledge base}. By prioritizing accessibility and scalability, the platform enables seamless integration into the daily routines of farmers, aligning with HCI best practices for AI systems designed for marginalized communities~\cite{amershi2019guidelines,shneiderman2020bridging}. 

Key features include:
\begin{enumerate}
 
   \item \textbf{Platform Accessibility:} Deployed on popular messaging apps like WhatsApp and Telegram, Farmer.Chat reduces the barriers to entry by leveraging familiar communication tools. This minimizes onboarding friction and fosters widespread adoption.
   \item \textbf{Multilingual and Multimodal Interaction: }The system supports Six languages Swahili, Amharic, Hausa, Hindi, Odiya, Telugu, and English, catering to diverse literacy levels. It further incorporates multimodal communication through text, voice notes, and images, enabling users with varying levels of literacy, particularly  users who find verbal or visual communication easier than text.
   \item  \textbf{Contextual Agricultural Knowledge Base:} The platform provides tailored advice across a variety of 40+ crops, including coffee, dairy, and potatoes, ensuring that information is relevant to local farming practices. Users receive personalized responses that align with their specific agricultural needs.
   \end{enumerate}
We will now briefly describe the interaction flow in Farmer.chat as shown in Figure~\ref{fig:interaction}. 
\begin{enumerate}
 
\item \textbf{Onboarding:} Farmer.Chat uses a structured, progressive onboarding process designed to ease the cognitive load on users with limited digital literacy. The bot introduces itself in simple terms, welcoming farmers to ask questions in local language or English, and reinforcing its role as a digital farming assistant. The stepwise flow—Start → Category → Crop Selection—ensures an intuitive introduction to the platform, based on lessons from human-centered research on rural tech adoption~\cite{medhi2006text,jain2018farmchat}.
\item	\textbf{User Interaction:} After onboarding, users choose a crop and ask specific questions. For instance, a user selecting ``coffee" and asking about the "benefits of the Ruiru variety" receives tailored, comprehensive information such as yield potential and disease resistance. Responses include voice notes in local languages for non-literate users, fostering a deeper engagement. The system encourages ongoing interaction by suggesting related topics and follow-up questions, promoting a personalized conversational experience (see Figure~\ref{fig:interaction}).
\item \textbf{Feedback Mechanism:} To maintain a feedback loop for continuous improvement, Farmer.Chat incorporates a two-tiered feedback system. Quick ratings via thumbs-up/down allow for immediate evaluation of response quality, while detailed star ratings capture more specific issues like incomplete or irrelevant content along with optional freeform feedback. These mechanisms not only enhance user engagement but also provide actionable data that informs iterative design refinements, ensuring the system remains responsive to evolving user needs.
\end{enumerate}

\begin{figure}[t!]
\centering
    \includegraphics[width=0.99\linewidth]{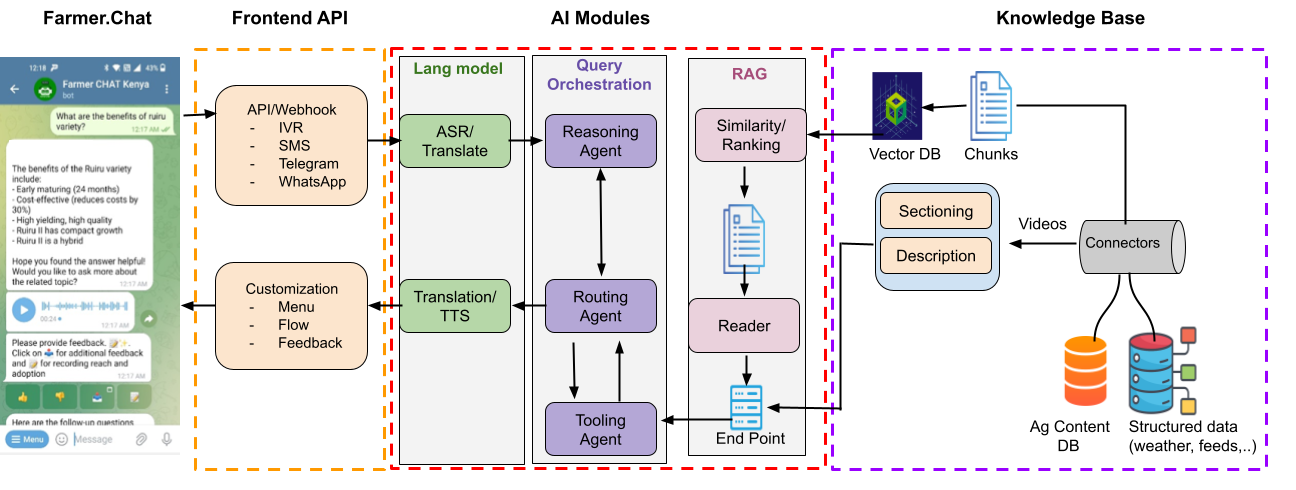}
    \vspace{-10pt}
    \caption{Overview of Farmer.CHAT Technical Architecture.}
    \label{fig:arch}
    \vspace{-15pt}
\end{figure}

\subsection{Farmer.CHAT Architecture: }
The architecture of Farmer.Chat is designed to ensure scalability, flexibility, and the delivery of accurate, contextually relevant agricultural advice as shown in Figure~\ref{fig:arch}. Its core consists of a robust \textit{Knowledge Base} and a suite of \textit{AI modules} that work together to process user queries and deliver tailored responses. This modular architecture is optimized for resource-constrained environments and supports both multimodal communication and real-time feedback integration.
\subsubsection{\textbf{Knowledge Base Builder: The Information Hub}}
\hfill\\
The Knowledge Base Builder is the foundation of Farmer.Chat, ingesting and organizing diverse data from numerous expert vetted sources, including research papers, policy documents, crop-specific guidelines, and multimedia content (e.g., images, videos). It also integrates agricultural knowledge on topics like banned chemicals, climate-smart farming, and local practices.

\begin{figure}[t!]
\centering
    \includegraphics[width=0.99\linewidth]{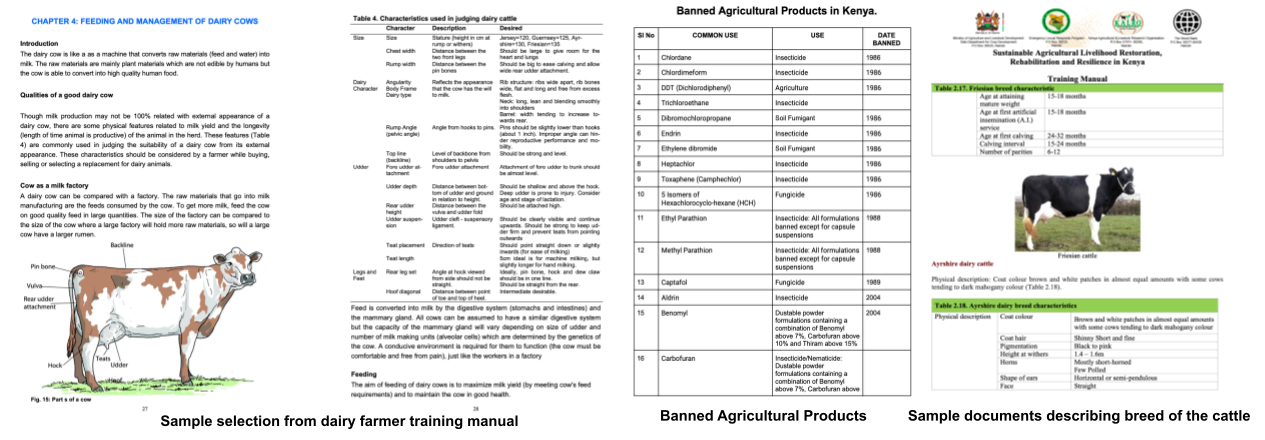}
    \vspace{-10pt}
    \caption{Diversity of Structured and Unstructured Documents Ingested in Farmer.Chat.}
    \label{fig:data}
    \vspace{-15pt}
\end{figure}

For instance, in the Dairy Value Chain in Kenya, the knowledge base includes: training manuals, gender study working paper, Climate-Smart Agriculture Practices Guide, Banned Agriculture Products Table, Dairy Housing Posters and Rural Sociology Papers. These documents vary in format, ranging from unstructured text (e.g., research papers) to highly structured tabular data (e.g., product catalogs), to multimedia files (e.g., posters and videos). Figure~\ref{fig:data} illustrates the diversity in Ag-related documents ingested in Farmer.Chat.

\textbf{a) Data Ingestion \& Processing:} The builder seamlessly pulls information from external sources including the web (government websites, agriculture universities documents, training manuals, policies, etc.), and YouTube (farming related videos). Through its Content Processing Engine, the system transforms unstructured data into structured formats using: Auto-summarization for fetching key insights, Categorization of content into agricultural topics, Auto-tagging for metadata management, Semantic chunking to group similar concepts for efficient retrieval.
The processed information is converted into vector embeddings for fast, context-aware search and retrieval, ensuring precise responses to user queries. The multimodal processing supports videos (via transcription, translation, and indexing) and images (via categorization, captions, and linking with agricultural practices).

\textbf{b) Scalability \& Flexibility:} The architecture allows easy expansion of new data, such as region-specific crops or climate data, as it becomes available. This ensures the system remains up-to-date and relevant for evolving agricultural needs.

\textbf{c) Feedback Loop:} The system incorporates user feedback into its continuous improvement cycle. Feedback from farmers—whether through thumbs-up/down or detailed star ratings—directly informs content prioritization and refinement, creating a dynamic, adaptive knowledge base.

\subsubsection{\textbf{AI Modules: Driving Intelligent Interactions}}
\hfill\\
The AI modules are the core of Farmer.Chat's interaction engine, ensuring real-time, contextually relevant responses to user queries. The architecture combines retrieval-based and generative capabilities, ensuring a conversational yet accurate experience for diverse literacy levels. The AI modules are composed of three key components:

\textbf{a)  Response Generator (RAG):} The Retrieval-Augmented Generation (RAG) approach underpins the system's response mechanism. It first retrieves relevant information from the vector knowledge base based on the user query and then uses a Large Language Model (LLM) to synthesize this into natural, easy-to-understand responses. This dual mechanism ensures the responses are both factually accurate and conversationally fluent, addressing users' diverse needs. By utilizing RAG, Farmer.Chat provides accurate, real-time responses that are deeply rooted in the knowledge base along with farmer-specific information (crop, location, and local weather, etc.) 

\textbf{b) Query Orchestration:} The Query Orchestration module is responsible for managing user interactions, simplifying complex queries, and ensuring coherent conversations. It comprises:

1.	\textit{Planning \& Reasoning Agent:} Identifies user intent and determines the most relevant data sources or tools for generating a personalized response.

2.	\textit{Execution Agent:} Retrieves, combines, or synthesizes data by interacting with APIs, the knowledge base, or external tools for real-time data, such as weather forecasts, pest information or market updates.

3.	\textit{Tooling Service:} Provides an extensible library of tools (e.g., real-time weather, pest diagnosis, or market data) that the LLM can access via APIs, ensuring the system remains adaptable as new agricultural technologies emerge.

\subsubsection{\textbf{Local Language, Voice \& Video Support: Enhancing Accessibility}}
\hfill\\
To cater to farmers with low literacy levels or language barriers, Farmer.Chat integrates comprehensive local language and voice support:

\textbf{a) Language Detection \& Translation:} The system automatically detects queries in multiple languages (e.g., Swahili, English) and translates them using Google Translate APIs~\cite{googletrans} for backend AI processing. Our evaluations show that generating responses directly in the source language results in poorer performance compared to translating the query into English, processing it via the LLM, and then back-translating the response—consistent with findings from recent multilingual LLM benchmarking efforts~\cite{ahuja2023mega}. This translation-based approach supports multilingual interactions and simplifies future language expansions.

\textbf{b) Speech Recognition and Text-to-Speech (TTS) Responses:} Farmers can use voice inputs, converted into text for the LLM, and receive audio replies by converting text~\cite{googletrans}. This multimodal interaction supports verbal communication, enhancing accessibility for non-literate users and those in rural areas where typing is less common.

\textbf{c) Video Ingestion and analysis:} Videos from sources like YouTube are ingested and their transcripts indexed. When a query is received, relevant videos and their transcripts are identified, and pertinent sections are used by Farmer.Chat to generate responses. This enables the system to leverage video content effectively to address user queries.

\subsubsection{\textbf{Integrating with Frontend Platforms: Seamless Interaction}}
\hfill\\
Farmer.Chat is deployed on popular platforms like WhatsApp and Telegram, leveraging their familiar interfaces to enhance adoption and usability. Its flexible API architecture allows for easy integration with other communication channels, such as IVR systems or dedicated mobile apps, ensuring future scalability.

Telegram Integration: The system leverages Telegram’s Bot API for real-time communication, enabling farmers to send and receive messages, voice notes, images, and other media. The API also supports structured menu navigation, simplifying complex interactions.

\subsubsection{\textbf{Continuous Learning: Feedback Loop for System Improvement}}
\hfill\\
A key strength of Farmer.Chat is its ability to evolve through a continuous feedback loop that integrates user interactions into the knowledge base, allowing for ongoing refinement and learning.

\textbf{a) Conversation Logs:} All user interactions are captured in detailed conversation logs, which are analyzed by human annotators to refine chatbot responses and improve system accuracy. These logs also help generate “golden” Q/A pairs, which allows us to retrain the AI models to enhance contextually appropriate responses.

\textbf{b) Contextual Adaptation:} Insights from conversation logs allow the system to adapt its responses to better align with user needs, making interactions more personalized and context-aware. 

\textbf{c) Content Improvement:} Analytics tools continuously scan for gaps in the knowledge base and also based on user interactions, ensuring that content remains relevant, accurate, and comprehensive.

\textbf{d) User Engagement:} Metrics on user interactions help evaluate how well the chatbot is meeting user needs, providing insights for enhancing the user experience.

\textbf{e) Query Effectiveness:} Performance metrics assess the accuracy and relevance of responses, identifying areas where the system can be optimized.

\textbf{f) Feedback Integration:} User feedback, both quantitative (e.g., satisfaction ratings) and qualitative (e.g., comments), is integrated into the system for iterative improvements in chatbot functionality and performance.

Through these mechanisms, Farmer.Chat continually adapts and evolves, ensuring that it remains responsive to the changing needs of its users while delivering accurate, timely, and relevant agricultural information. See Section~\ref{sec:quant} for more details.


\begin{table}[t!]
\caption{Deployment Details of Farmer.Chat Across Different Regions.}
\label{tab:dep}
\resizebox{\columnwidth}{!}{%
\begin{tabular}{|l|l|p{.6\textwidth}|l|l|l|}
\hline
Region &
  Content &
  Crops Covered &
  Other Services &
  No of Users &
  No of Queries \\ \hline
Kenya &
  Documents + Videos &
  Amaranth, Avocado, Banana, Banned Chemicals, Beans, Cabbage, Chicken, Coffee, Dairy, Maize, Mango, Onion, Pesticide Hazards, Potato, Poultry, Tomato &
  \begin{tabular}[c]{@{}l@{}}Weather \\ Disease diagnosis\end{tabular} &
  8805 &
  225500 \\ \hline
\begin{tabular}[c]{@{}l@{}}India   \\ (states of Bihar, UP, MP, \\ Rajasthan, Odisha, \\ Jharkhand, AP, Telangana)\end{tabular} &
  Documents + Videos &
  Banana,   Barley, Bio pesticides, Bitter gourd, Bottle gourd, Brinjal, Cabbage,   Cauliflower, Chickpea, Chilli, Coriander, Corn, Cucumber, Food processing and   value addition, French Bean, Green Gram, Guava, Lentil, Linseed (Flaxseed),   Long Beans, Maize, Mango, Mustard, Okra, Onion, Paddy, Papaya, Peas, Pigeon   Pea, Potato, Radish, Ridge gourd, Sesame, Spinach, Sponge gourd, Tomato,   Wheat &
  \begin{tabular}[c]{@{}l@{}}Weather \\ Disease diagnosis\end{tabular} &
  6282 &
  45798 \\ \hline
Nigeria &
  Documents &
  Maize,   Rice, Soybean &
  None &
  418 &
  3788 \\ \hline
Ethiopia &
  Documents + Videos &
  Agriprenuership,   Animal Feed, Animal Health, Apiculture (Beekeeping), Apple, Barley, Beans,   Beekeeping practices, Beetroot, Cabbage, Camels, Carrot, Cattle, Chickpea,   Coffee, Colony Establishment, Dairy, Donkey, Feed Management, Fertilizer   Application, Field Pea, Fruits, Garlic, Goat, Ground Nut, Haricot bean,   Herbicide Application, Hive management, Honey harvesting, Housing and   Environment, Irrigation, Land Use, Legumes, Len Seed, Lentile, Lettuce,   Maize, Meat production, Mechanization, Mung bean, Nipper Seed, Onion, Others,   Pastoralism, Pea, Potato, Poultry, Productivity, Rapeseed, River pump, Rope   Water Pump, Safflower, Sesame, Sheep, Soil Health Management, Soil Moisture,   Soil Nutrient Management, Sorghum, Soybean, Sunflower, Teff, Tomato, Vegetables,   Water Resource Management, Weeding, Wheat &
  None &
  105 &
  458 \\ \hline
\end{tabular}%
}
\end{table}

\section{Deployment and Implementation Details}
\label{sec:deployment}
Farmer.Chat has been deployed across multiple geographic regions to evaluate its scalability and adaptability to varying agricultural landscapes and user needs. The platform supports six languages: Swahili, Amharic, Hausa, Hindi, Odiya, Telugu, and English, alongside dialects like Bhojpuri. This diverse language support ensures accessibility across a broad spectrum of farmers, particularly in regions where literacy and language pose barriers to information access.

\textbf{a) Embeddings Model:} The system uses an custom embeddings model that can handle both structured and unstructured data, making it adaptable to region-specific data sets and user requirements.

\textbf{b) Deployment Channels:} The user interface is flexible, with deployment through channels like Telegram (operated as separate bots per region) and a standalone mobile app accessible globally.

Farmer.Chat’s deployment configurations are tailored to regional contexts, covering a wide array of crops and agricultural data. In addition to crop advisory, third-party services like weather forecasts from TomorrowIO~\cite{tomorrow} and disease diagnostics from Plantix~\cite{plantix} are integrated, enhancing the platform’s utility for farmers. These additional services are critical for region-specific agricultural decision-making, aligning with local farming practices.

To date, \textbf{15,000+ users} across all 4 user categories have used Farmer.Chat, handling \textbf{over 300,000 unique queries} in the past year, demonstrating the platform’s reach and engagement. Table~\ref{tab:dep} provides the deployment details of Farmer.Chat across 4 countries.

\subsection{Implementation: A Multi-Layered System}
The initial version of Farmer.Chat was launched on Telegram in Kenya in October 2023. The system has since undergone significant iterative improvements based on user feedback. Each iteration has focused on enhancing both the technical capabilities and user experience, ensuring that the system remains responsive to the needs of farmers (see Section~\ref{sec:quali} and~\ref{sec:design_impl} on user insights and design implications). The current version employs a multi-step LLM (Large Language Model) pipeline to process and generate responses efficiently:
\begin{enumerate}

\item 	\textbf{Intent Understanding: }The first step involves identifying user intent, which routes the query appropriately.
\item 	\textbf{Query Rephrasing \& Decomposition:} Farming-related questions are rephrased based on chat history and decomposed into smaller queries to facilitate precise retrieval.
\item \textbf{Text Retrieval \& Ranking:} The system retrieves relevant text snippets from the knowledge base, classifies them, ranks them based on relevance and prepares it for the final response.
\item 	\textbf{Response Generation:} Finally, the system generates a natural language response using an LLM, ensuring the language is both accurate and easy to understand. The system relies on a mix of GPT-4 for filtering and re-ranking and GPT-3.5 for other tasks. 
    
\end{enumerate}
For tasks involving translation, automatic speech recognition (ASR), and text-to-speech (TTS), the pipeline is configurable with Google Translate, Whisper, or equivalent systems, ensuring smooth language processing across diverse languages and dialects. We are committed to democratizing access to AI-driven agricultural support by making the Farmer.Chat codebase open to the global community. This will empower developers and researchers worldwide to build upon and adapt the platform across diverse regions and languages.

\begin{figure*}[!t]
\vspace{-5pt}
\begin{minipage}[l]{0.4\linewidth}
        \includegraphics[width=0.98\columnwidth]{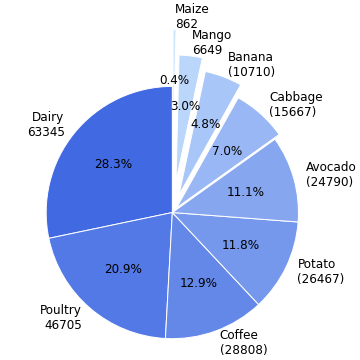}
        \subcaption{User Queries across Crops.}
        \label{fig:query_crop}
\end{minipage}
\begin{minipage}[l]{0.5\linewidth}
       \includegraphics[width=0.98\columnwidth]{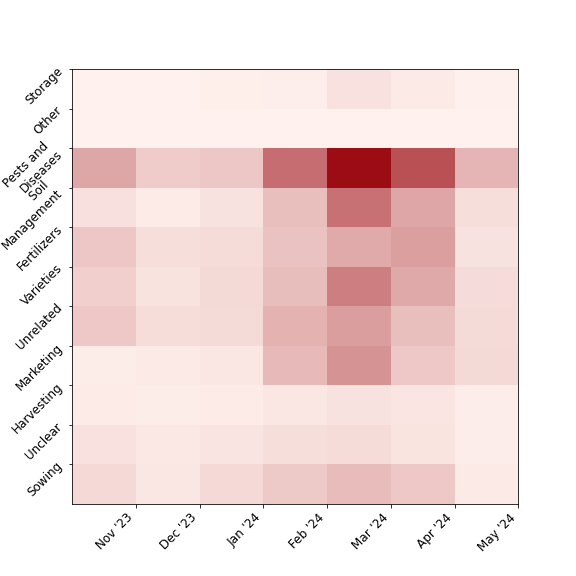}
       \subcaption{Topic Count Heatmap across Months.}
        \label{fig:topic_count}
\end{minipage}
\vspace{-5pt}
 \caption{Farmer.Chat User Queries and Engagement Patterns.}
        \label{fig:}
        \vspace{-18pt}
\end{figure*}

\subsection{Farmer.Chat Usage and Engagement Patterns }
This section focuses on the Farmer.Chat user engagement in Kenya, where it has seen the highest adoption rates. Farmer.Chat has been deployed across seven counties, serving an active user base of 8805 users, including 4729 men and 4076 women users. With a total of 225,500+ queries processed, users averaged 28.92 queries each, indicating a high demand for information.

Figure~\ref{fig:query_crop} shows that the most queried topics revolve around livestock, particularly Dairy (27.9\%) and Chicken (20.65\%). This suggests that livestock-related concerns are central to the users’ needs in Kenya. Additionally, crops like Potato (11.72\%), Avocado (10.93\%), and Coffee (12.63\%) received substantial attention, highlighting the importance of high-value crops to users. These insights underscore the necessity of expanding and refining content around these specific areas to better serve user demands.

Figure~\ref{fig:topic_count} illustrates monthly query patterns, revealing a significant surge in inquiries related to pests and diseases during the peak growing season in March'24 and April'24. This spike emphasizes the farmers' need for timely interventions, particularly for managing pest and disease outbreaks. Other frequently queried topics, such as Soil Management and Varieties, point to a growing need for educational content that provides farmers with localized, actionable strategies for soil nutrient management and selecting crop varieties best suited to their region.

To analyze user interactions in Farmer.Chat, we used a two-pronged approach. First, we performed a qualitative analysis through focused group discussions and direct user interactions to gain insights into farmers' engagement, challenges, and platform use during key agricultural periods. Second, we conducted a quantitative analysis with LLMs and machine learning tools to evaluate response quality, engagement, user query clarity, etc., to understand the performance and iteratively improve user experience. 

\section{Qualitative User Studies}
\label{sec:quali}
This section presents the qualitative research underpinning the design and iterative development of Farmer.Chat, a digital agricultural platform. Our primary goal was to understand how distinct user personas—Agriculture Extension Agents (AEAs), Agripreneurs, Lead Farmers, and general Farmers—interact with Farmer.Chat, providing insights to inform platform improvements and enhance user engagement. The research adopted a user-centered design approach, examining users' behaviors, challenges, and motivations through an intersectional lens. We aimed to:

\begin{enumerate}
    \item \textbf{Understand User Personas:} We explored users' daily realities, including their agricultural practices, pain points, and motivations, to ensure Farmer.Chat effectively serves their needs.
  \item \textbf{Evaluate User Engagement and Experience:} We assessed what drives user satisfaction and retention, focusing on usability, relevance of information, and overall user experience to identify areas for platform enhancement.

\end{enumerate}
The qualitative study began three months post-deployment, ensuring users had sufficient experience with the platform. Conducted across seven Kenyan counties, participants were selected based on active engagement with Farmer.Chat.

\textbf{Methodology:}
Participants, representing diverse farming roles and demographics, were recruited through a selection process ensuring inclusivity. Informed consent was obtained, with transparency regarding research objectives and time commitments~\footnote{This study was approved by the Institutional Review Board of the authors’ institution}. Data was gathered through in-depth interviews, usability tests, focus groups, and shadowing sessions. The research combined structured and open-ended questions to capture rich, nuanced insights into user experiences.

\textbf{Key Research Questions:}
\begin{enumerate}
    \item	What are the specific agricultural information needs of different personas (AEAs, Agripreneurs, Lead Farmers, and Farmers)?
  \item	How do users interact with Farmer.Chat to solve agricultural challenges, and what are the main use cases?
  \item	What motivates ongoing engagement, and how is Farmer.Chat perceived relative to traditional information sources?
  \item	What usability barriers do users encounter, and how can they be addressed in future iterations?
  \item	How can Farmer.Chat support gender equity and culturally sensitive information delivery?
\end{enumerate}

\subsection{Understanding User Personas}
To gain a comprehensive understanding of Farmer.Chat’s diverse user base, two user studies were conducted in \textbf{February} and \textbf{July 2024}. The research was guided by three main questions:

\begin{enumerate}
    \item	\textbf{Role, Responsibilities, and Digital Access:} How do the roles, responsibilities, and digital access of AEAs, Agripreneurs, Lead Farmers, and Farmers shape their interactions with Farmer.Chat?
    \item \textbf{Challenges and Motivations:} What specific challenges, motivations, and digital literacy barriers distinguish these user personas, particularly along the dimensions of age, gender, and education?
    \item	\textbf{Access to Trusted Information:} How can Farmer.Chat effectively address the need for reliable information, particularly for women farmers and those with limited digital literacy?
\end{enumerate}
\textbf{Methodology: In-Depth Interviews and Shadowing}

\noindent We employed in-depth interviews and shadowing to capture detailed personal narratives, focusing on digital access, challenges, and professional aspirations. 
\begin{itemize}
    \item \textbf{In-Depth Interviews:}
We developed semi-structured interview guides for each user segment. Interviews captured demographic data and explored key themes like access to agricultural information, digital literacy, and community roles. This method provided insights into how gender, age, and location influence users' needs and interactions with Farmer.Chat.

    \item \textbf{Shadowing:}
Shadowing sessions lasted six to eight hours, during which researchers observed participants’ work environments, social dynamics, and technology use. This method allowed for the collection of contextual data that enhanced the relevance and specificity of the interview questions.
\end{itemize}

\begin{figure}[t!]\centering
    \includegraphics[width=0.9\linewidth]{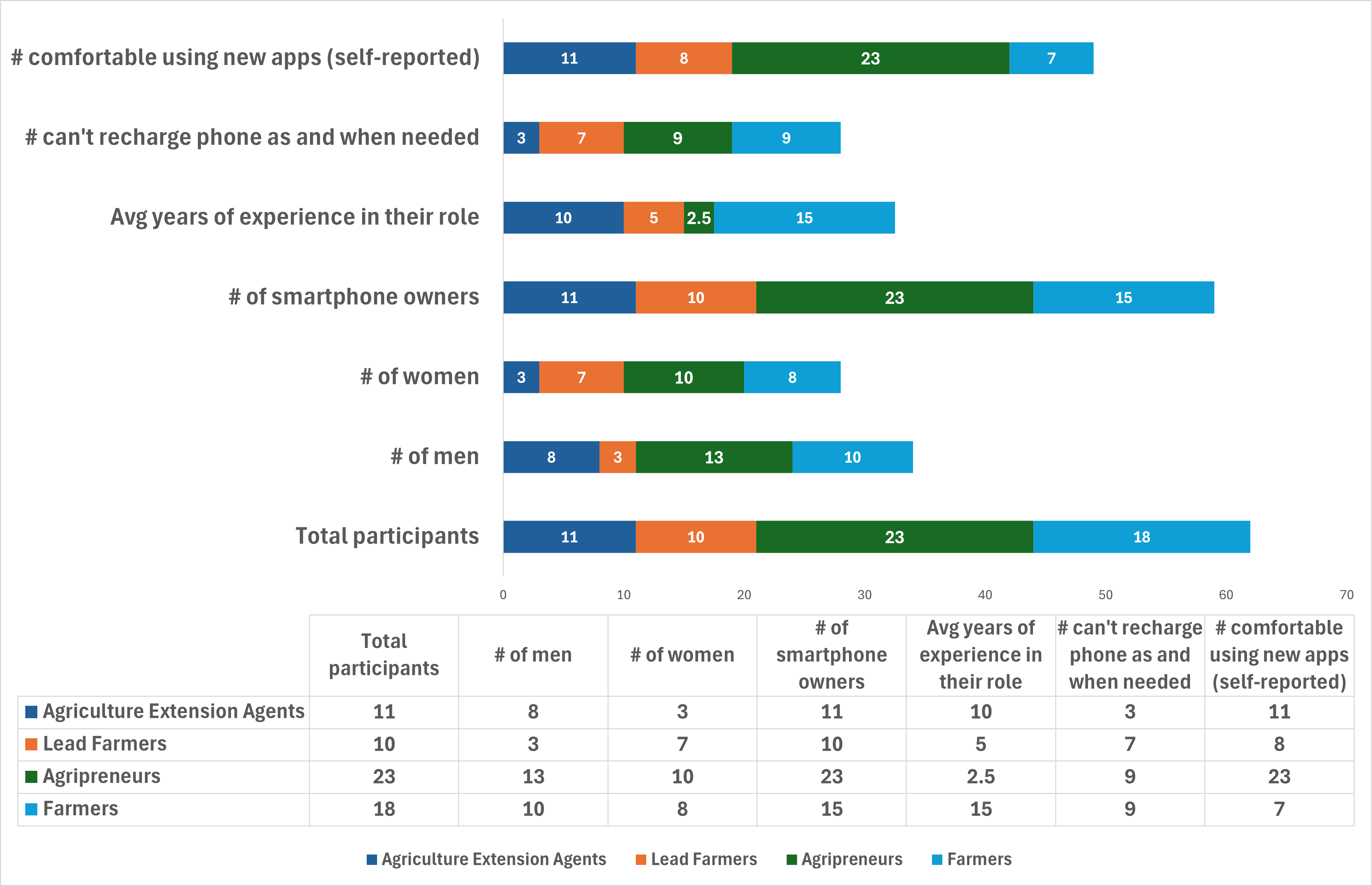}
    \vspace{-10pt}
    \caption{Demographics of Participants for the User Study on Understanding User personas.}
    \label{fig:demography}
    \vspace{-15pt}
\end{figure}
\noindent\textbf{Participant Selection:}
Figure~\ref{fig:demography} shows the demographics of the participants in this user study. To ensure diversity, participants were selected based on the following factors:

\textbf{•	Professional Role:} AEAs, Agripreneurs, Lead Farmers, and general Farmers.

\textbf{•	Gender:} Male and female participants were equally represented to account for gender-specific challenges.

\textbf{•	Age:} Participants ranged from 20 to 60+ years to capture generational differences in digital literacy.

\textbf{•	Geography:} Selected from seven counties to represent regional diversity in agricultural practices.

\textbf{•	Farm Type:} Farmers were further categorized by their primary value chain (e.g., dairy or mixed cropping), ensuring a diverse sample of farming practices.
\begin{figure}[t!]\centering
    \includegraphics[width=0.9\linewidth]{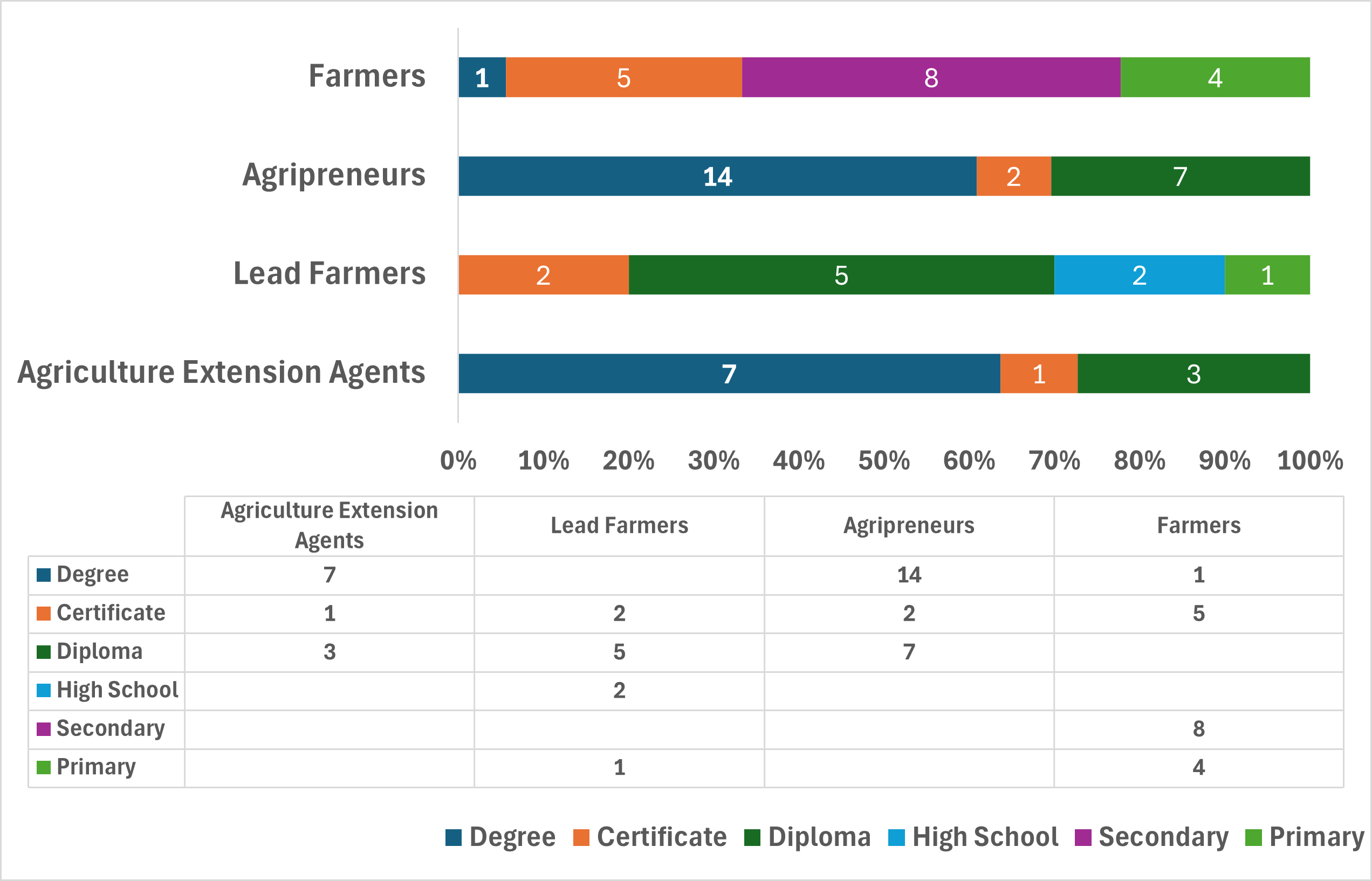}
    \caption{Education Background of the Participants in this User Study.}
    \label{fig:education}
    \vspace{-15pt}
\end{figure}

Figure~\ref{fig:education} shows the varying literacy levels across all user segments. In Kenyan education context, Primary is Grade 1-5, Secondary is grade 6-9, High school is grade 10-12. Certificate is after the high school level, and is a 6 month-1 year course, focusing on foundational knowledge. Diploma is higher than certificate, but lower than degree. It is 2-3 years course providing advanced training. Degree here means a Bachelor’s degree, which is 3-4 years long.

\subsubsection{\textbf{Finding 1: Demographics of User Segments}}
\hfill\\
We identified four distinct user segments in our research: Agriculture Extension Agents (AEAs), Agripreneurs, Lead Farmers, and Farmers. The demographic information collected helped us understand the diversity of professional roles, social responsibilities within their communities, frequency of interaction with farmers, and access to smartphones. 

\textbf{\textit{Key Insights:}}
\begin{itemize}
    \item \textbf{Age and Gender Distribution:} Significant variability was observed in age and gender across the user segments. AEAs, Lead Farmers, and Farmers presented a wide age range, while Agripreneurs were predominantly under 35. Gender representation was more balanced in AEAs and Lead Farmers, but women were underrepresented among Farmers and Agripreneurs.
    \item \textbf{Smartphone Access:} Nearly all AEAs, Agripreneurs, and Lead Farmers have access to smartphones and use them regularly. However, some farmers, especially older individuals and women, reported not owning smartphones.
    \item \textbf{Data Affordability:} Despite widespread smartphone access, data affordability was a major issue for Lead Farmers and Farmers, particularly in rural areas. While AEAs and Agripreneurs had access to data, they were constrained in their ability to use digital tools regularly due to the high cost of mobile data.
\end{itemize}

\subsubsection{\textbf{Finding 2: Criteria of User Personas.}}
\hfill\\
We conducted a thematic analysis to develop user personas as they serve as archetypes that represent groups of users with similar needs, pain points, motivations, and behaviors.

\textbf{\textit{Key Criteria for Personas:}}

\begin{itemize}
    \item \textbf{Leadership:} Within AEAs, the extent to which they take leadership in disseminating information varies greatly. More active AEAs engage frequently with farmers, while others do so less consistently.
\item \textbf{Digital Literacy:} Across segments, digital literacy emerged as a major differentiator, with younger Agripreneurs and AEAs typically demonstrating higher competence in using digital tools. Older Lead Farmers and Farmers—particularly women—tended to have lower levels of digital literacy.
\item \textbf{Motivations:} Motivations varied across personas. Some users, especially AEAs and Lead Farmers, were motivated by a desire to serve their communities and be role models. Others, particularly Agripreneurs, were driven by the goal of generating income and progressing in their careers.
\item \textbf{Age and Education:} Among Lead Farmers and Farmers, age and educational background were significant differentiators. Older users often had less formal education, which directly affected their comfort with technology and access to agricultural information.
\end{itemize}
\subsubsection{\textbf{Finding 3: Pain Points of Users}}
\hfill\\
Across all user segments, several common pain points emerged, primarily centered around access to trusted, timely agricultural information.

\textbf{\textit{Key Pain Points:}}
\begin{itemize}
    \item \textbf{Limited Access to Trusted Information:} A universal pain point for all personas was the difficulty in accessing reliable, validated agricultural information when needed. This was especially acute for women farmers, who reported that existing channels of advisory services often failed to address their specific needs.
\item \textbf{Outdated or Incomplete Information:} AEAs reported that they frequently received farmer queries on topics beyond their technical expertise, leading to delays in providing accurate advice. Farmers and Lead Farmers struggled to find reliable information across the agricultural value chain—from seed selection to production and marketing. This included critical topics like weather patterns and government subsidies.
\item \textbf{Challenges with Digital Tools:} Users with smartphones, particularly younger Agripreneurs and tech-savvy AEAs, often resorted to using online search engines or platforms like YouTube. However, they found it challenging to determine whether the information was credible and endorsed by agriculture experts or government bodies.
\item \textbf{Gender and Access Gaps:} For women users, these pain points were further exacerbated by limited access to digital tools and trusted networks. As a result, they experienced delays in accessing timely agricultural advice, which directly impacted their productivity and economic outcomes.
\item \textbf{Value Proposition of Farmer.Chat:} The ability to provide verified, expert-endorsed agricultural information via Farmer.Chat emerged as a strong value proposition across all user segments. Users recognized the potential for Farmer.Chat to fill the gap in providing trustworthy and timely advice.
\end{itemize}
\subsection{Understanding User Experience }
User experience is a multifaceted concept encompassing users’ interactions with Farmer.Chat, including their thoughts, feelings, and perceptions. Our study aimed to explore the experiences of Farmer.Chat users, focusing on three key aspects: the \textbf{usefulness} of Farmer.Chat in their work lives, the \textbf{trustworthiness and quality} of the responses they received, and the \textbf{ease of interacting} with the chatbot. Key research questions: 
\begin{enumerate}
\item How do users perceive the usefulness of Farmer.Chat in addressing their agricultural needs, and what value does it bring to their work?
\item What are users' levels of trust in the responses provided by Farmer.Chat, and how do they evaluate the quality and accuracy of information?
\item How easy or difficult do users find interacting with Farmer.Chat, and what specific usability challenges do they face?
\item What differences exist in the user experience of various user segments (e.g., gender, frequency of use)?
\end{enumerate}
\noindent\textbf{Methodology: Focus Group Discussions, In-depth Interviews, and Usability tests}

\begin{wrapfigure}{r}{5cm}
\vspace{-10pt}
    \includegraphics[width=0.33\columnwidth]{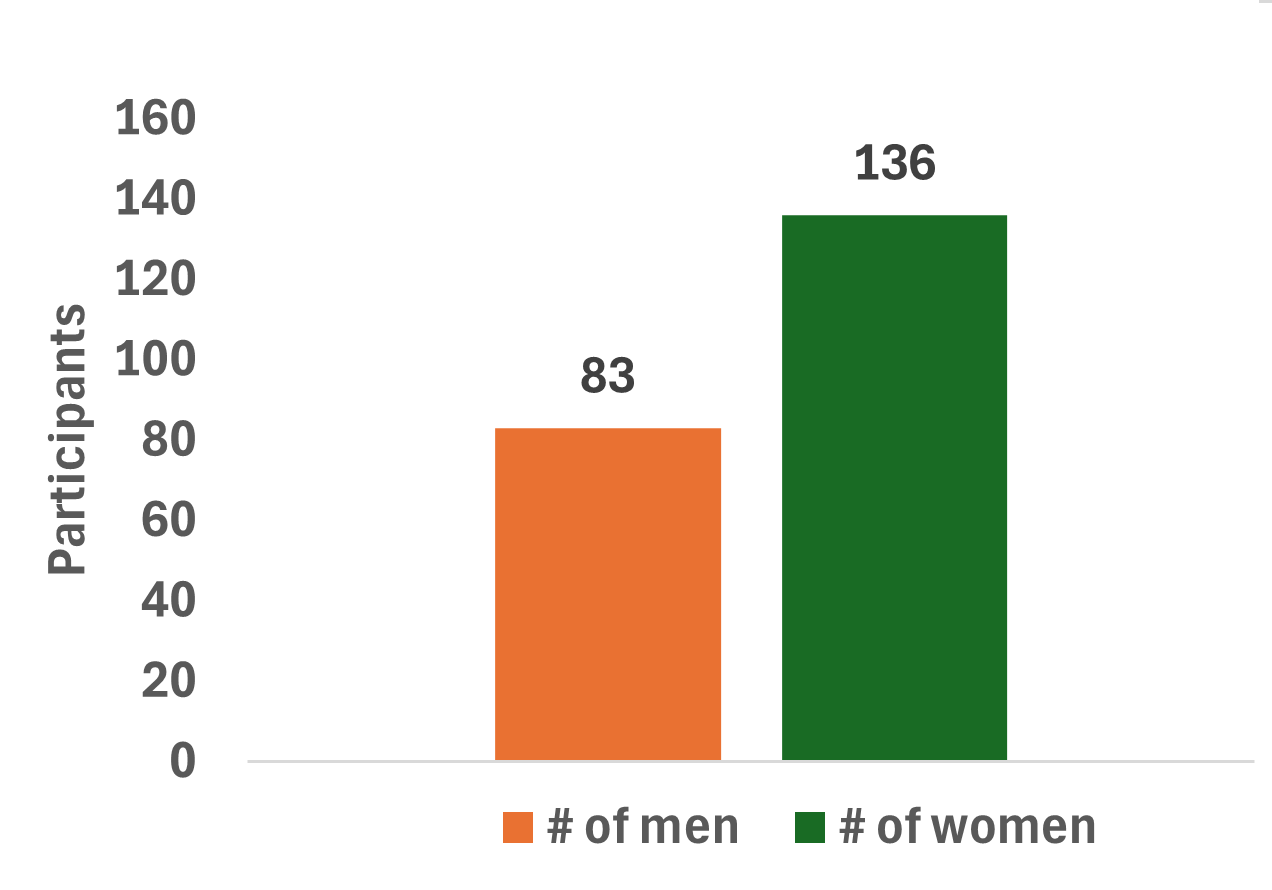}
    \vspace{-15pt}
    \caption{Study Participants Details.}
    \label{app:fig:inference}
    \vspace{-15pt}
\end{wrapfigure}
Two user studies were conducted to gather feedback on Farmer.Chat. The first, in \textbf{January 2024}, involved active users who had been interacting with the chatbot regularly, defined as those submitting at least two queries per week. The second study, in \textbf{April 2024}, specifically focused on women lead farmers to understand gendered experiences and potential barriers.

To gather comprehensive feedback, we employed focus group discussions, in-depth interviews, and usability tests. This mixed-methods approach allowed us to examine both the social and technical dimensions of user experience.
\begin{itemize}
  
   \item \textbf{Focus Group Discussions:} Focus group discussions (FGDs) enabled dynamic interactions among users, facilitating rich, qualitative insights into their experiences. The discussions explored common themes affecting user experience, such as the reasons for using Farmer.Chat and the challenges encountered. FGDs were held in two counties, Nyeri and Meru, in January 2024, involving 199 users (116 Female, 83 Male) through 14 group discussions. In April 2024, additional FGDs were held with 20 women lead farmers in Uasin Gishu county to explore gender-specific experiences and barriers in greater depth.

   \item \textbf{In-Depth Interviews:} Following the FGDs, we conducted in-depth interviews to delve deeper into individual experiences. These interviews were held with seven selected users (three men, four women) to explore how frequent and infrequent users differed in their interactions with Farmer.Chat. The selection criteria were based on how often they engaged with the bot, providing a diverse range of user experiences.

   \item \textbf{Usability Tests:} Usability testing was performed to observe how users interacted with Farmer.Chat and to assess any difficulties they encountered. The observations were held with seven selected users (three men, four women). During these tests, users demonstrated how they used the chatbot on their smartphones, showing real examples of their interactions. The test focused on key features, question formats (text or voice), and how users navigated the platform.
\end{itemize}
\subsubsection{\textbf{Finding  1:  Usefulness in Accessing Information.}}
\hfill\\
Active users with leadership roles, either formal or informal, who are also technically proficient with smartphones and motivated to learn, reported the most value from Farmer.Chat. These users, often interacting with multiple farmers daily, found the bot particularly helpful in addressing frequent farmer queries that they couldn't answer immediately. Before using Farmer.Chat, they spent considerable time and resources searching online, verifying information, or consulting with supervisors. Now, Farmer.Chat offers immediate, trusted responses, significantly enhancing their work efficiency and confidence. This not only boosted their engagement with farmers but also led to greater job satisfaction. 
A Lead Farmer shared their experience:
\begin{displayquote}
“I want to see farmers do well, and that is my main motivation. Agriculture Extension Officers can't be reached easily by individual farmers. If they form farmer groups, then Agriculture Extension Officers address the entire group. So I formed a farmers' group in my community. At times, I call Agriculture Extension Officers , who are not always available or don't answer the phone. So that problem is solved now with Farmer.Chat.”  \textbf{Lead Farmer, 52yrs, M, in in-depth interview.}
\end{displayquote}

\subsubsection{\textbf{ Finding 2: Value Proposition and Practical Benefits.}}
\hfill\\
Users who experienced tangible benefits, such as improved crop yield or effective control of pests and diseases, expressed strong satisfaction with Farmer.Chat. A recurring theme among users was cost-saving, where users were able to find solutions directly from the bot instead of hiring professionals. For example:
\begin{displayquote}
“My wife suggested I check out FarmerChat. We have been considering adding avocados to our farm but we were not sure where to find reliable growing information. The bot arrived just when we needed it. I have been using it primarily for avocado-related queries and have learned a lot about the crop's diseases and pests.” \textbf{Farmer, 43 yrs, M, Nakuru county, FGD.}

“I use the bot when there is a challenge, like a disease in the livestock. My cow had lumpy skin disease. I asked the bot and got a solution. It saved the consultation fee that I would have needed to pay the vet. It also saved time since I did not have to wait for the vet or to get back to me.” \textbf{Lead Farmer, 30-40 yrs, F, Meru County, FGD.}
\end{displayquote}
In addition, the platform served as a valuable learning tool, with users appreciating the ease of asking multiple questions—something they hesitated to do with human advisors.

\subsubsection{\textbf{Finding 3: Building Trust in Farmer.Chat.}}
\hfill\\
For agriculture extension agents and lead farmers who disseminate information to others, trust in Farmer.Chat was key. Trust often developed after users tested the advice on a small scale in their own fields. Seeing positive results strengthened their confidence in the platform, making them more likely to share the advice with others. However, some users remained skeptical, especially when the advice differed from their previous knowledge. Trust in the bot was significantly boosted when users knew the information came from verified sources, such as the Ministry of Agriculture or other trusted sources.
\begin{displayquote}

“Once I found that whatever advice had been given for me was working in my own farm, then I shared with other farmers, who asked what I had done differently.” \textbf{Lead Farmer, 42 Yrs, M, Meru county, in a FGD. }
    
\end{displayquote}

\subsubsection{\textbf{ Finding 4: Quality of Responses.}}
\hfill\\
The quality of responses—defined as relevance, detail, and understandability—was a critical factor for user satisfaction. Most users found the responses contextually appropriate, though some desired more actionable details, such as the availability of inputs required for recommended practices. A key pain point was the language used; many users, including those with technical expertise, found that overly technical or complex measurements made it hard to apply the advice. Respondents requested that the bot use simpler, conversational language to ensure clarity, especially for less educated farmers.
\begin{displayquote}
“Use of simple english would be helpful. Don’t use a lot of mathematics in the response, there are farmers who didn’t go to school, and wouldn’t be able to understand this.” \textbf{Lead Farmer, 55yrs, F, Nyeri county, in in-depth interview.}
\end{displayquote}

\subsubsection{\textbf{Finding 5: Ease of Interaction.}}
\hfill\\
Ease of interaction was a major differentiator in user engagement. More tech-savvy users, familiar with the chatbot’s features and limitations, were able to articulate their queries effectively and continue to engage with Farmer.Chat regularly. 
\begin{displayquote}
“I started using the bot immediately after onboarding. I asked any questions that I could think of. When you search using a keyword, like in [a search engine], you get 10 answers, which are saying different things. But here on the bot, you get the right answers. The answers are 100\% correct. I always get the right answer.” \textbf{Lead Farmer, 30yrs, F, Meru county, in an in-depth interview.}
\end{displayquote}
In contrast, users with lower digital literacy struggled to phrase their questions in a way the bot could understand, leading to frustration and reduced engagement. 
\begin{displayquote}
“Simplify the menu for the old farmers like my mother [so that they can use it easily].” \textbf{Lead Farmer, F, Nyeri county, during qualitative survey.}

\end{displayquote}

\subsection{Continuous User Feedback through Qualitative Surveys.}
To gather ongoing feedback on Farmer.Chat's usability and overall user experience, we recently started conducting continuous qualitative studies through bi-weekly phone surveys starting in August 2024. These surveys captured real-time user experiences without imposing significant time or travel burdens on rural users.

\textbf{Research Focus:} The qualitative surveys aimed to explore three key areas:
\begin{enumerate}
 \item 	\textbf{User Satisfaction: }How satisfied were users with their overall experience?
 \item 	\textbf{Feature Evaluation:} What aspects of the bot did users find most useful or problematic?
 \item \textbf{Improvement Suggestions:} What recommendations did users have for enhancing the bot’s functionality and content?
\end{enumerate}

\noindent\textbf{Methodology:}
We conducted two rounds of qualitative phone surveys with 20 participants each, balancing gender (10 male, 10 female) and representing seven counties across rural Kenya. 
Participants were selected based on two criteria: (1) users who had asked at least one question on the chatbot in the past month, and (2) an equal gender split, ensuring diverse perspectives across seven counties in rural Kenya. Each survey lasted 5–10 minutes and collected both open-ended and Likert-scale responses 
All participants provided informed consent prior to participation. We also ensured that participation was voluntary, with users able to withdraw at any time.

\begin{figure}[t!]\centering
    \includegraphics[width=0.65\linewidth]{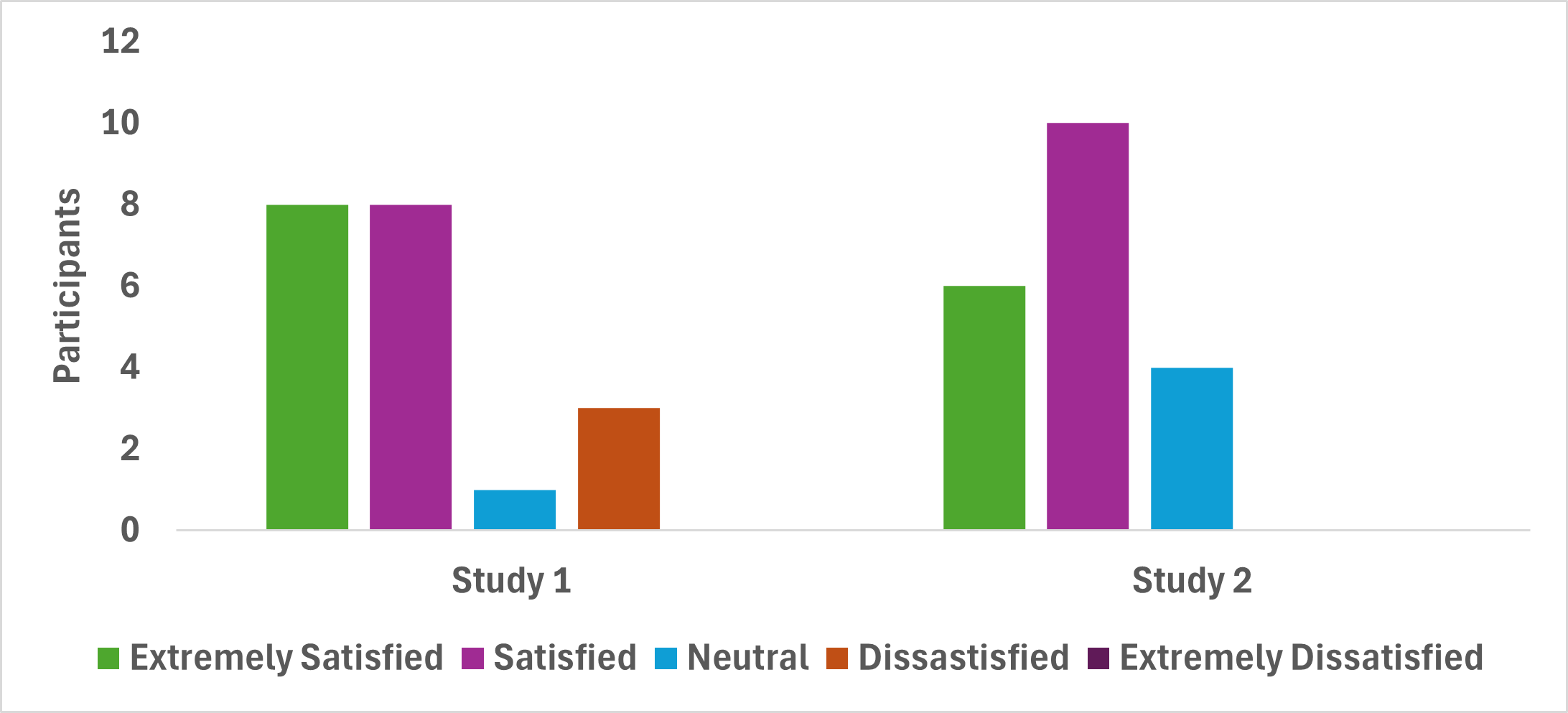}
    \vspace{-10pt}
    \caption{Participants Feedback on Farmer.Chat Overall Experience.}
    \label{fig:ongoing_feedback}
    \vspace{-15pt}
\end{figure}
Figure~\ref{fig:ongoing_feedback} shows the participants feedback from both the studies on how satisfied they were with the overall experience in Farmer.Chat.

\textbf{\textit{Study 1 Results:}}

•	\textbf{Satisfaction:} Most users were satisfied, with 8 extremely satisfied, 8 satisfied, 1 neutral, and 3 dissatisfied.

•	\textbf{Practical Benefits:} 5 users noted the bot’s role in improving yields, managing pests, and acting as a reliable resource.

•	\textbf{Ease of Use:} 12 users found the bot easy to navigate, though some desired more detailed responses.

•	\textbf{Challenges:} A few users struggled with the interface, highlighting the need for more training.

•	\textbf{Suggestions:} Users requested expanded content (e.g., market linkages, crop calendars), training, offline mode, and local language support in navigating the menu.

\textbf{\textit{Study 2 Results:}}

•	\textbf{Satisfaction:} 8 were extremely satisfied, 10 satisfied, and 4 neutral.

•	\textbf{Impact:} Users successfully addressed farming issues through the bot, particularly with livestock, poultry, and new crop varieties.

•	\textbf{Reduced Usage:} Some users reduced usage due to external factors (e.g., relocation, education).

•	\textbf{Suggestions:} Calls for more content, guaranteed responses, technical improvements, and continued offline support.

\subsection{Design Implications and Iterative Improvements of Farmer.Chat}
\label{sec:design_impl}
\begin{figure}[t!]\centering
    \includegraphics[width=0.9\linewidth]{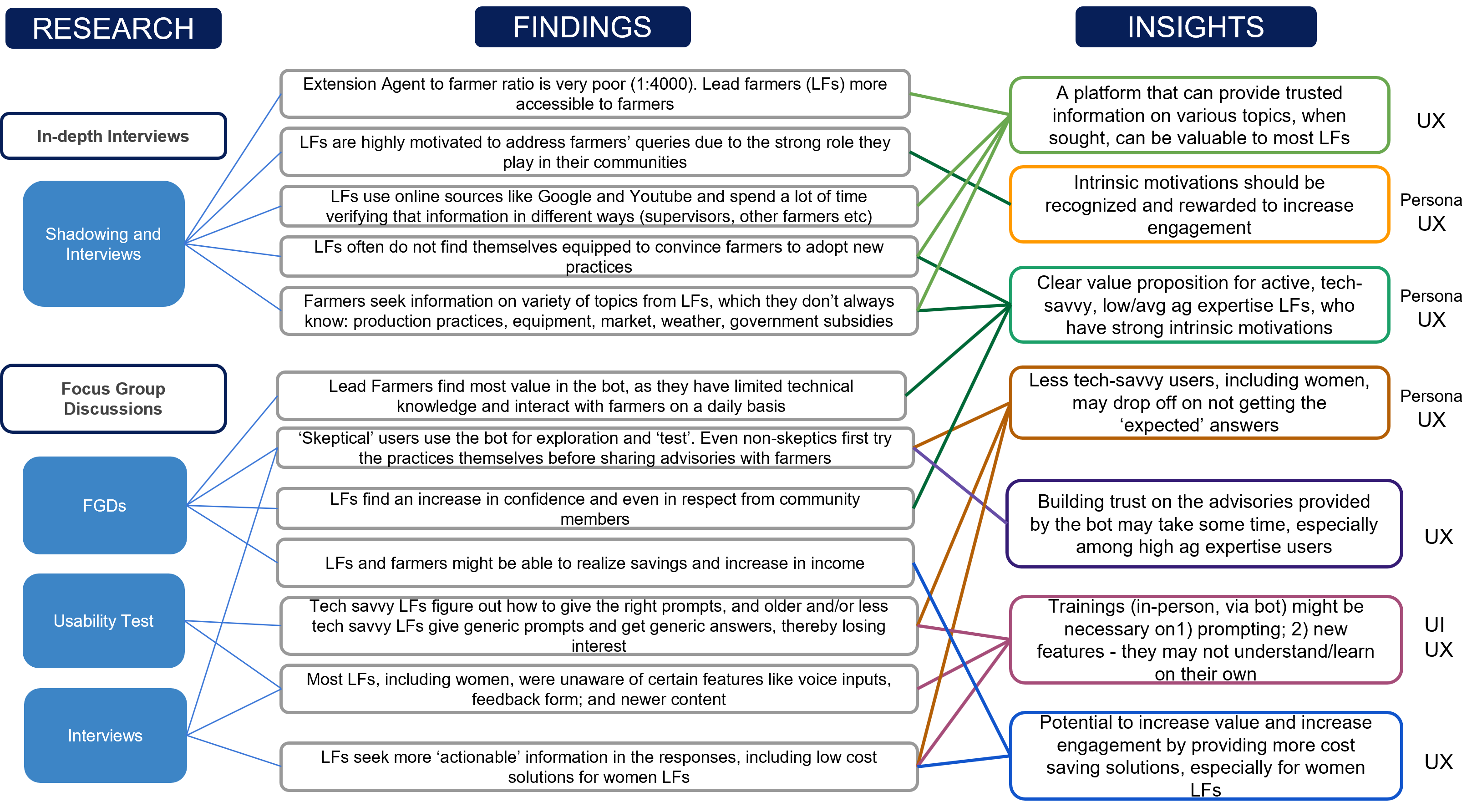}
    \vspace{-10pt}
    \caption{Farmer.Chat Design Implications: User Study Research to Findings to Insights.}
    \label{fig:impl1}
    \vspace{-15pt}
\end{figure}

\begin{figure}[t!]\centering
    \includegraphics[width=0.9\linewidth]{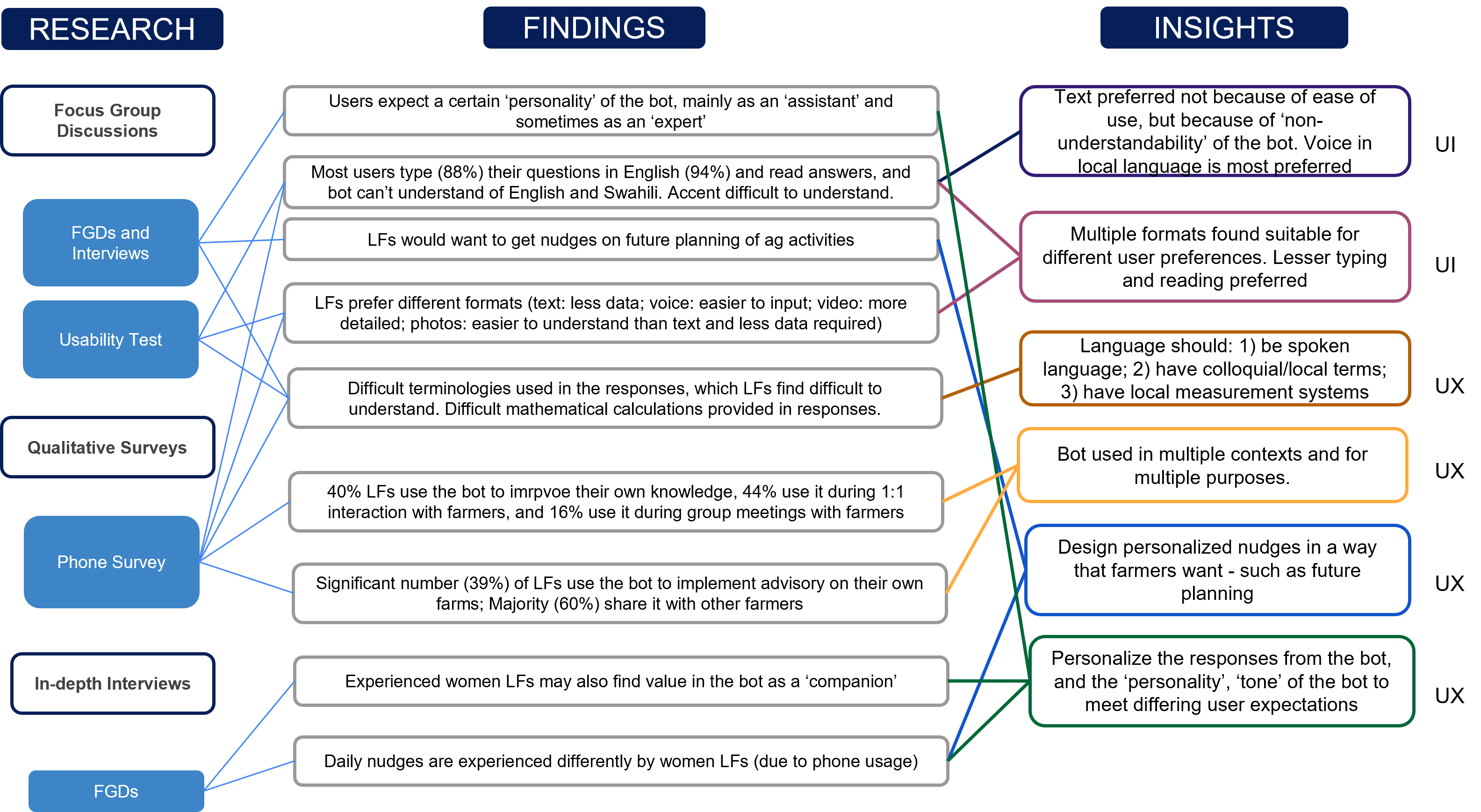}
    \vspace{-10pt}
    \caption{Farmer.Chat Design Implications: User Study Research to Findings to Insights.}
    \label{fig:impl2}
    \vspace{-15pt}
\end{figure}

Based on our extensive user studies and continuous feedback, several key design implications emerged, leading to significant improvements in Farmer.Chat. Figure~\ref{fig:impl1}~and ~\ref{fig:impl2} shows the summarized design implications resulting from user research studies. 
\begin{itemize}

    \item \textbf{Trusted, Expert-Endorsed Information:} Trust remains a key factor, especially for farmers seeking reliable advice. Farmer.Chat now prioritizes content verification and endorsements from credible authorities, such as the Ministry of Agriculture, to enhance the platform's credibility.

    \item \textbf{Improved Information Retrieval and Response Efficiency:} To meet the needs of users requiring fast, reliable data, the system has been optimized for quicker content retrieval. Features like follow-up questions enable users to explore topics more deeply without typing new queries, boosting engagement, as shown in usage analytics (see Quantitative Analysis section).

    \item \textbf{Simplified Interface for Diverse User Base: }Acknowledging the varying technical proficiency of users, Farmer.Chat's interface has been simplified for clearer navigation. Voice input support and starter prompts tailored to users' locations ensure easy access to relevant information, improving overall usability for less tech-savvy users.

    \item \textbf{Enhanced Training and Support through Video Tutorials:} To support users less familiar with digital tools, short 20-30 second video tutorials were introduced, guiding users on accessing new features. These tutorials, shared through online groups used by AEAs and Lead Farmers, have significantly boosted feature adoption.

    \item 	\textbf{Expanding Practical Content and Features:} Based on user feedback, Farmer.Chat has broadened its content to include resources like market linkages, crop calendars, and localized agricultural insights. This continuous expansion ensures users receive actionable and detailed guidance.
\end{itemize}

\section{Results: Quantitative Analysis }
\label{sec:quant}
This section presents a detailed quantitative analysis of Farmer.Chat, categorized into four key areas: User-centric metrics, Insight-centric metrics, Accuracy metrics, and Responsible AI metrics. 



\subsection{User-Centric metrics}
In this section, we focus on key user-centric metrics to assess the usability, accessibility, and overall user experience in Farmer.Chat. These metrics include the complexity of responses, the percentage of unanswered questions, and latency, which together offer insights into how well Farmer.Chat serves its diverse user base. 

\begin{figure*}[!t]
\vspace{-5pt}
\begin{minipage}[l]{0.4\linewidth}
        \includegraphics[width=0.98\columnwidth]{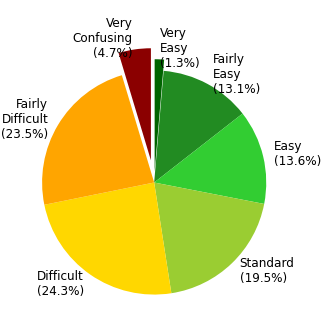}
        \subcaption{Difficulty Level Distribution of all Responses using FK Score.}
        \label{fig:diff_level}
\end{minipage}
\begin{minipage}[l]{0.4\linewidth}
       \includegraphics[width=0.98\columnwidth]{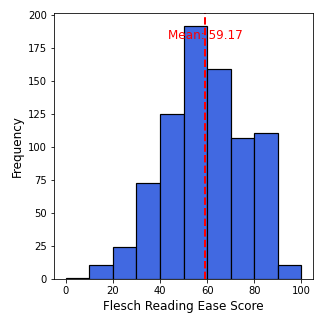}
       \subcaption{Flesch-Kincaid (FK) Score Distribution.}
        \label{fig:fkscore}
\end{minipage}
\vspace{-5pt}
 \caption{Analysis on Complexity of Responses.}
        \label{fig:}
        \vspace{-18pt}
\end{figure*}
\subsubsection{\textbf{Complexity of Responses}}
\hfill\\
This metric evaluates how easily users, with varying literacy levels, comprehend Farmer.Chat's information. Agricultural content often includes technical jargon, making it essential to balance simplicity with maintaining critical details. While overly complex responses can overwhelm users, oversimplified ones may lose value.

We use established readability metrics like the Flesch-Kincaid (FK) score~\cite{fkscore,textstat}, which assesses the ease of reading based on sentence length and syllable count, ranging from 0 to 100 (higher scores indicate easier text). Farmer.Chat responses typically score between 60 and 80 (See Figure~\ref{fig:diff_level} and~\ref{fig:fkscore}). While this suggests overall accessibility, certain responses remain slightly more complex than ideal for some users.

While readability scores like Flesch-Kincaid provide valuable insights, they need to be benchmarked and possibly adjusted for agriculture-specific contexts. For example, a term that may be considered ``difficult" in a general setting may be commonly understood by farmers but still challenging for others. Our qualitative findings align with this, showing that users find Farmer.Chat response highly valuable and comprehensible, with strong satisfaction expressed in our in-depth interviews and focus group discussions. 

\begin{figure*}[!t]
\vspace{-5pt}
\begin{minipage}[l]{0.4\linewidth}
        \includegraphics[width=0.98\columnwidth]{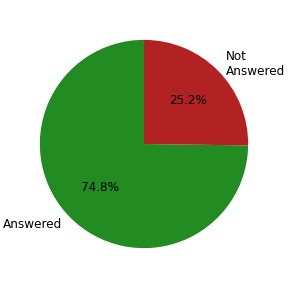}
        \vspace{-10pt}
        \subcaption{Percentage of Queries that received a Response.}
        \label{fig:answer}
\end{minipage}
\begin{minipage}[l]{0.4\linewidth}
       \includegraphics[width=0.98\columnwidth]{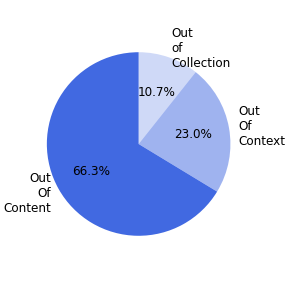}
       \vspace{-10pt}
       \subcaption{Percentage of Unanswered Query Categories.}
        \label{fig:unanswer}
\end{minipage}
\vspace{-5pt}
 \caption{Analysis of Percentage of Answered and Unanswered Questions.}
        \label{fig:answer_analsyis}
        \vspace{-18pt}
\end{figure*}
\subsubsection{\textbf{Percentage of Unanswered Questions}}
\hfill\\
This metric provides valuable insights into Farmer.Chat's ability to respond effectively to user queries. As illustrated in Figure~\ref{fig:answer}, the platform successfully answers close to 75\% of user questions. However, analyzing the reasons for unanswered queries is critical for identifying areas of improvement.

We categorized unanswered questions using an LLM (GPT4) into three key types:\\
•	\textbf{Out of Context:} Queries unrelated to agriculture, falling outside the platform's scope.\\
•	\textbf{Out of Collection:} Queries about unsupported crops or value chains.\\
•	\textbf{Out of Content:} Queries where the knowledge base lacks sufficient information.

Our analysis shows that 66\% of unanswered queries are due to content gaps (Out of Content), 23\% are Out of Context, and 11\% are Out of Collection (see Figure~\ref{fig:unanswer}).

To address these gaps, we are continuously updating the knowledge base, reducing the frequency of Out of Content queries, and expanding coverage to provide more comprehensive support across diverse agricultural topics.
\begin{wrapfigure}{r}{5cm}
    \includegraphics[width=0.3\textwidth]{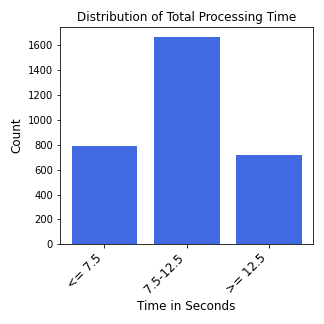}
    \vspace{-10pt}
    \caption{Response Time Distribution.}
    \label{fig:latency}
    \vspace{-18pt}
\end{wrapfigure}
\vspace{-5pt}
\subsubsection{\textbf{Response Time}}
\hfill\\
Response time is critical for user satisfaction in real-time applications like Farmer.Chat. The average response time in Farmer.Chat is 9.05 seconds, but this distribution skews right due to a few queries requiring more context during retrieval. The majority of the time is spent on reranking (3.15 seconds) and generation (2.71 seconds). This is justified as these steps are crucial for ensuring that the most contextually relevant and accurate responses are delivered to users. Figure~\ref{fig:latency} shows that most user queries (68\%) fall within the 7.5-12.5 seconds range. Only 15\% of queries take more than 12.5 seconds to process, mainly due to complex retrieval scenarios.

In a domain-specific application like Farmer.Chat, where providing highly accurate agricultural advice is paramount, a latency of 5-10 seconds is generally acceptable. Importantly, none of the users in our studies expressed concerns about the response time, as their focus remained on the quality and relevance of the information provided.

\subsection{Insight-Centric Metrics }
Insight-Centric mtrics offer valuable insights into user interactions with Farmer.Chat, focusing on User Behavior, Content Utilization, and Query/Prompt Clarity to improve performance and user experience.

\begin{figure*}[!t]
\begin{minipage}[l]{0.6\linewidth}
        \includegraphics[width=0.98\columnwidth]{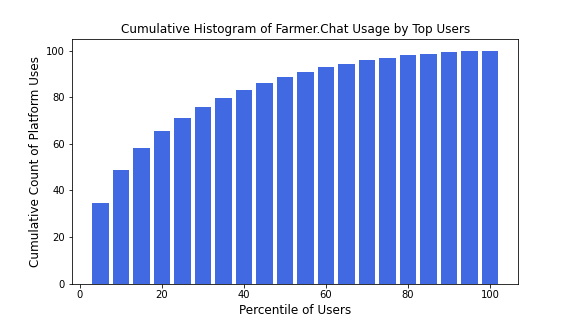}
        \subcaption{Cumulative Percentage of Queries on Farmer.Chat by its Users.}
        \label{fig:cumu}
\end{minipage}
\begin{minipage}[l]{0.35\linewidth}
       \includegraphics[width=1.5\columnwidth]{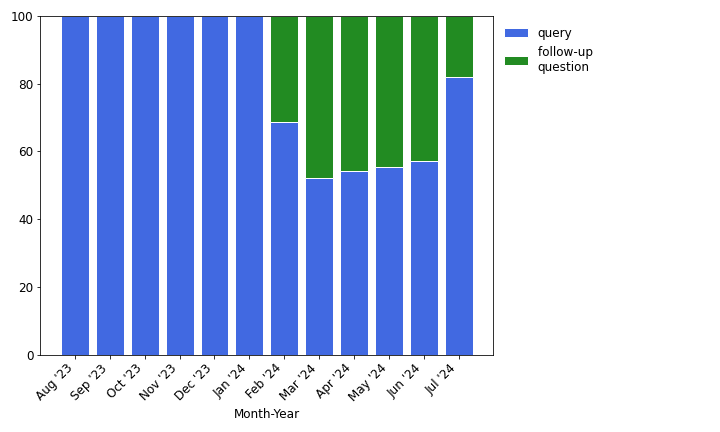}
       \subcaption{User Queries vs Follow up Queries}
        \label{fig:follow}
\end{minipage}
\vspace{-5pt}
 \caption{User Behavior Analysis.}
        \label{fig:}
        \vspace{-10pt}
\end{figure*}
\subsubsection{\textbf{User Behavior Analysis}}
\hfill\\
We focused on understanding user engagement patterns, specifically analyzing the frequency of questions asked by different user segments. Our analysis shows that 35\% of users, termed \textbf{``power users,}" contribute to nearly 80\% of total queries, making them key to driving engagement and retention on the platform (Figure~\ref{fig:cumu}).

Our qualitative user studies revealed that many users struggled to formulate follow-up questions, leading most to ask only one question per topic. To address this issue, in February'24, we introduced a feature that provides easily clickable follow-up questions after an initial response. This feature was designed to help users refine or expand their queries, which are often incomplete or unclear in the agricultural context. As illustrated in Figure~\ref{fig:follow}, this addition had a significant impact, contributing to over 45\% of total interactions during peak periods. 

\begin{figure}[t!]\centering
    \includegraphics[width=0.6\linewidth]{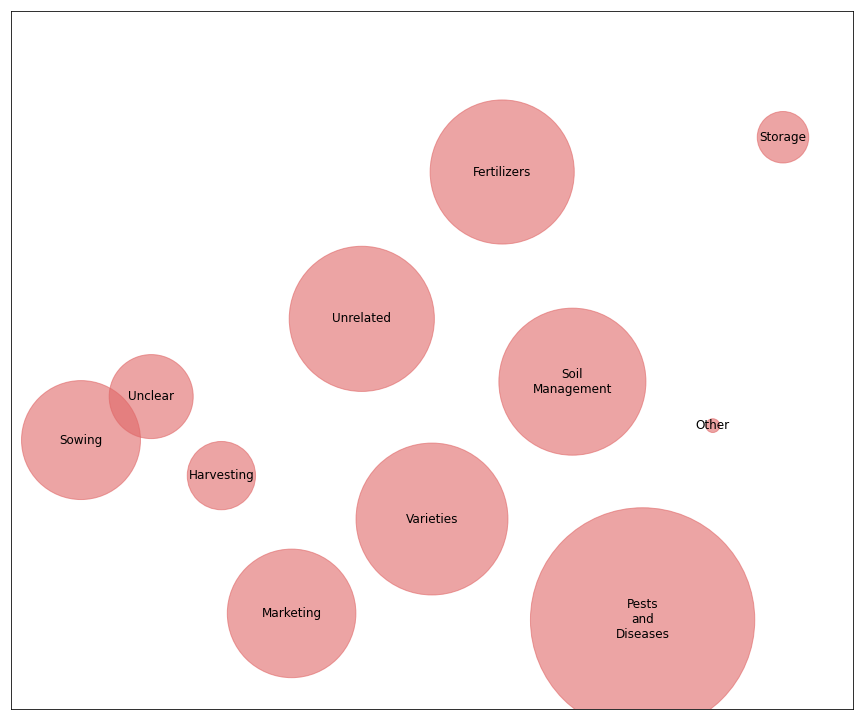}
    \caption{Bubble Chart Visualization of the Most Discussed Topics on Farmer.Chat.}
    \label{fig:topic-cloud}
\end{figure}

\begin{figure*}[!t]
\begin{minipage}[l]{0.4\linewidth}
        \includegraphics[width=0.98\columnwidth]{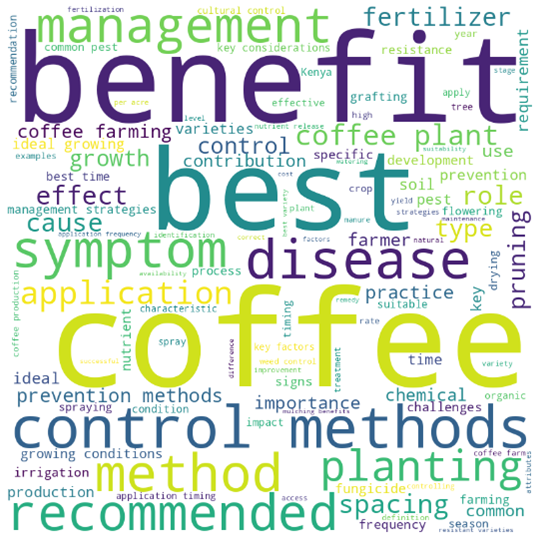}
        \subcaption{Most Frequently Mentioned Attributes for Coffee.}
        \label{fig:cloud1}
\end{minipage}
\begin{minipage}[l]{0.4\linewidth}
       \includegraphics[width=0.98\columnwidth]{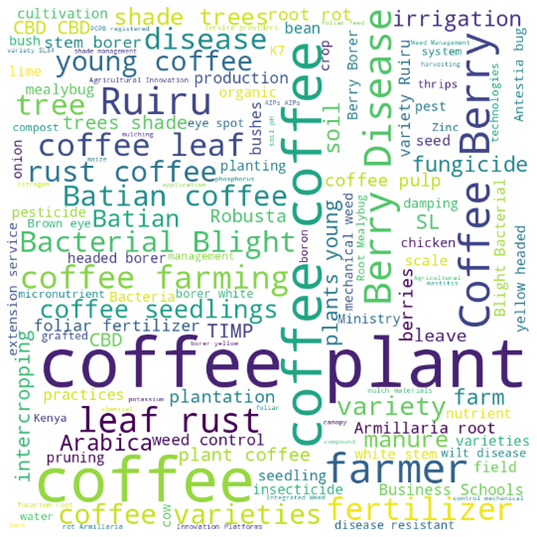}
       \subcaption{Most Frequently Mentioned Entities related to Coffee.}
        \label{fig:cloud2}
\end{minipage}
\vspace{-5pt}
 \caption{Word Cloud on Most Frequently Mentioned Attributes and Entities on Farmer.Chat.}
        \label{fig:word}
        \vspace{-18pt}
\end{figure*}
\subsubsection{\textbf{User Query Analysis: Most discussed Topics}}
\hfill\\
Figure~\ref{fig:topic-cloud} presents a bubble chart that visualizes the most discussed topics on the platform. The size of each bubble corresponds to the frequency of a topic's occurrence, with "Pests and Diseases" emerging as the most frequently queried area, followed by "Soil Management" and "Storage". These topics reflect the primary concerns of users, particularly around pest and disease control, optimizing soil health, and proper storage practices—key challenges in agriculture.

Figure~\ref{fig:cloud1} and~\ref{fig:cloud2} present word clouds highlighting key attributes and entities within the coffee value chain. Prominent attributes such as "disease," "rust," "borer," "fertilizer," "plantation," and "pruning" underscore a high demand for advice on disease management and cultivation techniques. The dominant entities, including "control," "planting," "management," and "growth," reflect users’ focus on effective disease control and best practices for planting and growth.

   
\begin{figure}[t!]\centering
    \includegraphics[width=0.6\linewidth]{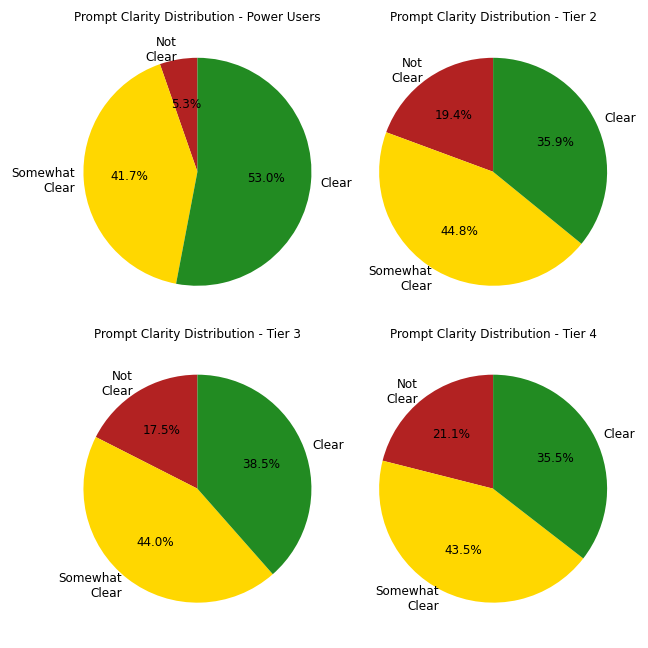}
    \caption{Query Articulation Clarity across User Segments.}
    \label{fig:query-art}
   
\end{figure}

\subsubsection{\textbf{User Query Analysis: Query articulation clarity}}
\hfill\\
To assess how effectively users articulate their queries, we developed a User Query/Prompt Clarity Score, evaluated by an LLM (GPT-4) based on three criteria: (a) score-1: clarity of intent, (b) score-2: topic specificity, and (c) score-3: entity-attribute identification. Scores range from 1 (unclear) to 3 (very clear), with the final score being the average of these criteria. The system prompt guiding the evaluation process is as follows:

\textit{“Given a user input about agriculture, score the prompt on a scale of 1 to 3 for each of the following criteria: (a) whether the intent is clear, (b) whether the prompt addresses a specific topic, and (c) whether the query references a specific entity and its attribute. Provide a reason for each score and calculate the final score as the rounded average of the three.”}

For instance, the input "Tell the benefits of Batian coffee variety" would score 3 for clarity of intent (clear ask), 3 for topic specificity (focused on coffee), and 2 for entity-attribute clarity (references Batian, but no specific attribute). This results in a final score of 3 (rounded). Table~\ref{tab:intent-ex} provides additional examples.

This analysis was conducted on over 20,000 user prompts, categorized by user activity levels into four tiers: Tier 1 (Top 5\%), Tier 2 (5-10\%), Tier 3 (10-20\%), and Tier 4 (remaining users). Tier 1 contained around 10,000 prompts, while the other tiers had between 3,000 and 4,000 each, ensuring a robust sample size across segments.

Figure~\ref{fig:query-art} shows the clarity score distribution for power users (Tier 1), revealing that 53\% of their prompts were classified as "Clear," 41.7\% as "Somewhat Clear," and 5.3\% as "Not Clear." This indicates that more active users articulate queries with greater precision, likely due to their familiarity with the platform.

\begin{table}[]
\caption{Table Illustrates Examples of Different User Queries and their Scoring for Query Clarity.}
\label{tab:intent-ex}
\renewcommand{\arraystretch}{2.8}
\resizebox{\columnwidth}{!}{%
\begin{tabular}{|p{.3\textwidth}|l|p{.3\textwidth}|l|p{.3\textwidth}|l|p{.3\textwidth}|l|}
\hline
User query &
  Score 1 &
  Reason &
  Score 2 &
  Reason &
  Score 3 &
  Reason &
  Total score \\ \hline
When   is the time to plant corn? &
  3 &
  The intent is   clear as the user is asking for information about the timing of planting   corn. &
  3 &
  The topic is   specific as it pertains to corn, which is a well-defined agricultural crop. &
  2 &
  While the input   is about corn, it does not specify an attribute such as a particular variety   of corn or specific conditions for planting. &
  3 \\ \hline
What   is the recommended fertilization schedule for early crop areas in coffee   farming? &
  3 &
  The topic is   specific as it pertains to coffee farming, particularly in early crop areas. &
  3 &
  The input is   focused on a specific entity (fertilization schedule) and its attribute   (recommendation for early crop areas in coffee farming). &
  3 &
  The input is   focused on a specific entity (fertilization schedule) and its attribute   (recommendation for early crop areas in coffee farming). &
  3 \\ \hline
How   does irrigation in coffee farming increase rainfall and benefit the plants? &
  2 &
  The intent is   somewhat clear as the user is asking about the relationship between   irrigation and rainfall in coffee farming, but the phrasing could be more   direct in specifying what benefits they are looking for. &
  3 &
  The topic is   specific to coffee farming, which is a well-defined area within agriculture. &
  2 &
  The input   mentions coffee farming but does not specify a particular entity or attribute   of the plants being discussed, such as growth rates or yield improvements. &
  2 \\ \hline
Coffee   spraying &
  1 &
  The intent is   unclear; it does not specify what information is being sought regarding   coffee spraying. &
  1 &
  The topic is   very broad and lacks specificity; 'coffee spraying' could refer to various   practices or issues. &
  1 &
  There is no   specific entity or attribute mentioned; it does not identify a particular   aspect of coffee spraying. &
  1 \\ \hline
\end{tabular}%
}
\end{table}

For the remaining tiers, the majority of prompts (43.5\% to 44.8\%) were rated as “Somewhat Clear,” with "Clear" prompts being the second most common. "Not Clear" prompts remained the smallest category across all tiers. This trend suggests that while clarity improves with user engagement, many users still face challenges in articulating clear queries.

\begin{figure*}[!t]
\begin{minipage}[l]{0.45\linewidth}
        \includegraphics[width=0.98\columnwidth, height=2in]{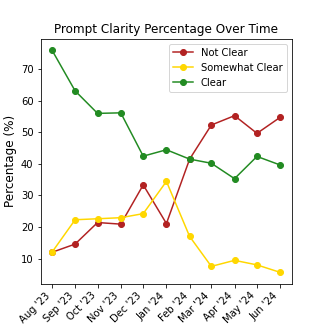}
        \subcaption{Query Clarity Over Time Across Users.}
        \label{fig:query-clarity}
\end{minipage}
\begin{minipage}[l]{0.45\linewidth}
       \includegraphics[width=0.98\columnwidth, height=2in]{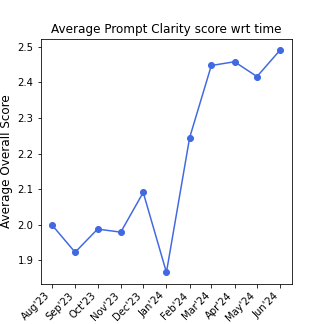}
       \subcaption{Overall Query Clarity Score Over Time}
        \label{fig:query-score}
\end{minipage}
\vspace{-5pt}
 \caption{Query Clarity Over Time.}
        \label{fig:}
        \vspace{-18pt}
\end{figure*}
\subsubsection{\textbf{User Query Analysis: Query Clarity Over Time}}
\hfill\\
Figure~\ref{fig:query-clarity} illustrates the progression of prompt clarity over time, categorized into "Clear," "Somewhat Clear," and "Not Clear." Initially, "Somewhat Clear" queries dominate, but over time, "Clear" queries gradually increase, indicating that users are becoming more skilled in articulating precise queries. The "Not Clear" category remains consistently low and declines further, suggesting continuous improvement in user query formulation as they gain familiarity with the platform. However, there's still room to reduce "Not Clear" queries to enhance the overall user experience.

Figure~\ref{fig:query-score} depicts the average prompt clarity score over time, with minor fluctuations around 2.25 on a 1 to 3 scale. This improvement is primarily driven by power users, who become more adept at interacting with Farmer.Chat through repeated use, demonstrating a self-learning process. This adaptive learning trend highlights the platform's ability to foster better natural language understanding over time.

Based on both user studies and quantitative analysis, we've integrated prompt suggestions to guide users as they begin their journey on Farmer.Chat, further improving query formulation.

\begin{figure}[t!]\centering
    \includegraphics[width=0.7\linewidth]{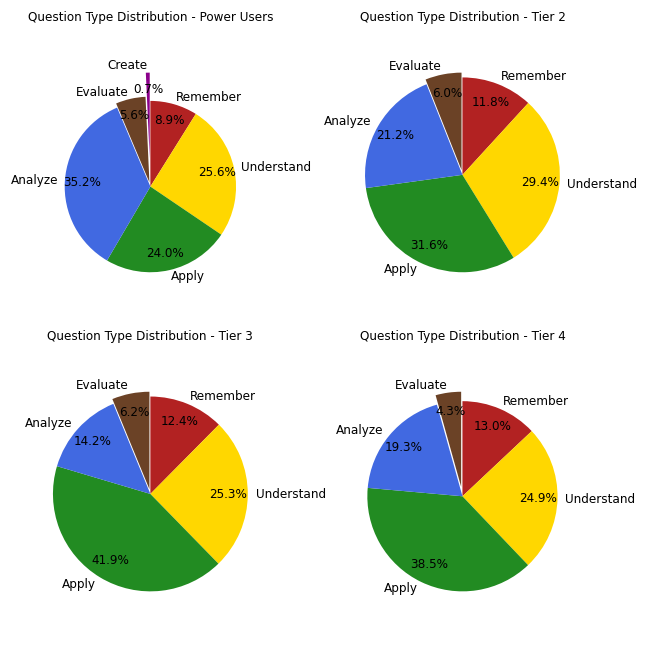}
    \vspace{-20pt}
    \caption{Cognitive Demands on the LLM based on User Queries.}
    \label{fig:bloom}
    \vspace{-22pt}
\end{figure}

\subsubsection{\textbf{Cognitive Ability Required by LLMs (Farmer.chat) to Process the User Query}}
\hfill\\
We assessed Farmer.Chat's cognitive demands on the LLM using Bloom's Revised Taxonomy~\cite{bloom1,bloom2}, which classifies cognitive processes into six levels: Remember, Understand, Apply, Analyze, Evaluate, and Create. Each user query was categorized into these levels to gauge the system's cognitive engagement requirements.

To evaluate the complexity of user queries, we prompted the LLM to classify each query into one of the six levels. Queries from four user tiers, based on engagement levels, were analyzed:

•	\textbf{Power Users (Top 5\% Users):} Figure~\ref{fig:bloom} shows that 32.6\% of queries fell into "Analyze," followed by "Understand" (27.3\%) and "Apply" (25.6\%). These users often engage in higher-order cognitive tasks, seeking to analyze, evaluate, and apply expert knowledge to solve complex agricultural issues.

•	\textbf{5-10\% Percentile Users (Tier-2):} Figure~\ref{fig:bloom} shows a focus on "Apply" (34.3\%), followed by "Understand" (25.6\%) and "Analyze" (24.6\%). These users prioritize applying knowledge to practical contexts but also engage in understanding and analyzing data for better decision-making.

•	\textbf{10-20\% Percentile Users (Tier-3):}  As seen in Figure~\ref{fig:bloom}, 45.5\% of queries from this group were classified as "Apply," indicating a primary focus on implementing actionable advice with less emphasis on understanding (21.4\%) and analyzing (14\%).

•	\textbf{Beyond the Top 20\% Users (Tier-4):} For users in the remaining 80\%, Figure~\ref{fig:bloom} reveals that "Apply" (40.1\%) remained dominant, followed by "Understand" (23.3\%) and "Remember" (13.2\%). These users mostly seek practical, easy-to-apply solutions with minimal cognitive engagement in understanding or analyzing.

•	\textbf{Cognitive Trends Across User Tiers:} Across all user tiers, there is a clear trend: as user engagement decreases, so does the cognitive complexity of their queries. Power users engage more with analysis and application, while lower-engagement users focus primarily on remembering and applying information.

\begin{figure}[t!]\centering
    \includegraphics[width=0.4\linewidth]{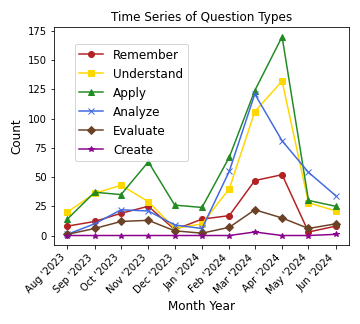}
    \caption{Distribution of Different Question Types across Varying Months.}
    \vspace{-10pt}
    \label{fig:question-type}
   
\end{figure}

\subsubsection{\textbf{Distribution of Question Types Across Different Months}}
\hfill\\
Figure~\ref{fig:question-type} illustrates the evolving distribution of question types on Farmer.Chat over several months, reflecting how users' needs change with the agricultural cycle. Table~\ref{tab:query-type} shows different user queries and their corresponding mapping to various query types.

\textbf{\textit{Key Findings:}}\\
\textbf{•	Remember:} Queries for retrieving specific information remain steady, with fluctuations that correspond to overall platform activity.\\
\textbf{•	Understand:} Questions aimed at comprehending agricultural concepts rise from October to April, peaking in April, as users prepare for the peak farming season.\\
\textbf{•	Apply:} Demand for practical, actionable advice also peaks in April, highlighting the need for guidance during key crop periods.\\
\textbf{•	Analyze:} Deep analytical queries spike in March, indicating farmers’ engagement in pre-season planning and evaluation.\\
\textbf{•	Create:} This category remains low, showing users' preference for established practices over innovative solutions.
\begin{table}[]
\caption{Illustrations of User Queries being Classified into Different Query Types.}
\label{tab:query-type}
\resizebox{\columnwidth}{!}{%
\begin{tabular}{|l|l|}
\hline
\multicolumn{1}{|c|}{\textbf{User query}}                                              & \multicolumn{1}{c|}{\textbf{Query type classification}} \\ \hline
When is the time to plant corn? & Remember \\ \hline
What is the recommended fertilization schedule for early crop areas in coffee farming? & Apply                                                   \\ \hline
How does irrigation in coffee farming increase rainfall and benefit the plants?        & Analyze                                                 \\ \hline
Coffee spraying                 & None     \\ \hline
What are the important periods of watering coffee plants during the wet season?        & Understand                                              \\ \hline
How can farmers decide when to water coffee plants based on humidity levels?           & Apply                                                   \\ \hline
\end{tabular}%
}
\end{table}
\textbf{\textit{Insights and Implications:}} The alignment between query types and the agricultural calendar suggests that users’ needs shift from basic information retrieval to practical application and critical planning as the season progresses. The early spike in "Analyze" queries emphasizes the importance of providing timely, relevant content that supports farmers throughout different phases of the crop cycle.

\subsection{Accuracy Metrics:}
Farmer.Chat uses a Retrieval-Augmented Generation (RAG) system to provide precise and contextually relevant responses. The evaluation of RAG's effectiveness focuses on two core areas: context retrieval and response quality.

\subsubsection{\textbf{Context Retrieval Metrics: Context Precision}}
\hfill\\
This metric assesses the proportion of relevant content retrieved in response to a user query. High precision ensures that irrelevant information is minimized in the retrieved text snippets, improving the overall accuracy and utility of responses. Inspired by the RAGAS library~\cite{es2023ragas}, precision is calculated by comparing retrieved passages against the user queries and scoring the relevance of those passages. This is especially important in agricultural contexts, where specific, nuanced information is often required.

For instance, if a user queries "symptoms and treatment for foot and mouth disease in cows," the system retrieves the top 10 text passages based on cosine similarity or a similar function. These passages are used to generate factual statements via an LLM, and each statement is rated as relevant (1) or irrelevant (0) to the query. The context precision score is the ratio of relevant statements to the total number of statements, reflecting the percentage of retrieved information that aligns with the input query.

In our evaluation of over 1,000 queries using different configurations of chunking, embeddings, and LLMs, we found that the average context precision in Farmer.Chat is 71\%.

\begin{figure}[t!]\centering
    \includegraphics[width=0.8\linewidth]{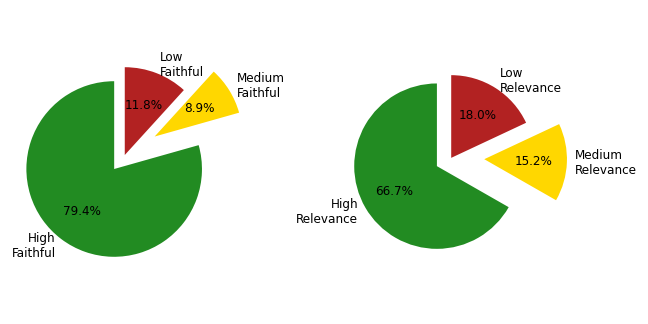}
    \vspace{-25pt}
    \caption{Response Quality Metrics- Faithfulness and Relevance.}
    \label{fig:faith}
   \vspace{-15pt}
\end{figure}

\subsubsection{	\textbf{Response Quality Metrics}}
\hfill\\
We evaluate response quality using two key metrics—faithfulness and relevance—through the RAGAS library~\cite{es2023ragas}. RAGAS utilizes an LLM to break down responses into factual statements and evaluate these metrics based on the retrieved context. Studies confirm that RAGAS metrics align closely with human evaluations, a finding we validated with an internal analysis of 1,000 queries from Farmer.Chat.
\begin{itemize}
    \item \textbf{Response Faithfulness:} This metric measures the factual accuracy of each statement within a response, comparing it to the context retrieved. Scores range from 0 (inaccurate) to 1 (accurate). For example, a query about bacterial blight misinterpreted as coffee late blight would yield a faithfulness score of 0.
    \item \textbf{Response Relevance:} This metric assesses how well the response addresses the user's query. A response only partially related to the original question would result in a lower relevance score.
\end{itemize}
Responses are categorized based on these scores: High Accuracy: Scores above 0.7, Medium Accuracy: Scores between 0.3 and 0.7, Low Accuracy: Scores below 0.3.

\begin{figure}[t!]\centering
    \includegraphics[width=0.9\linewidth]{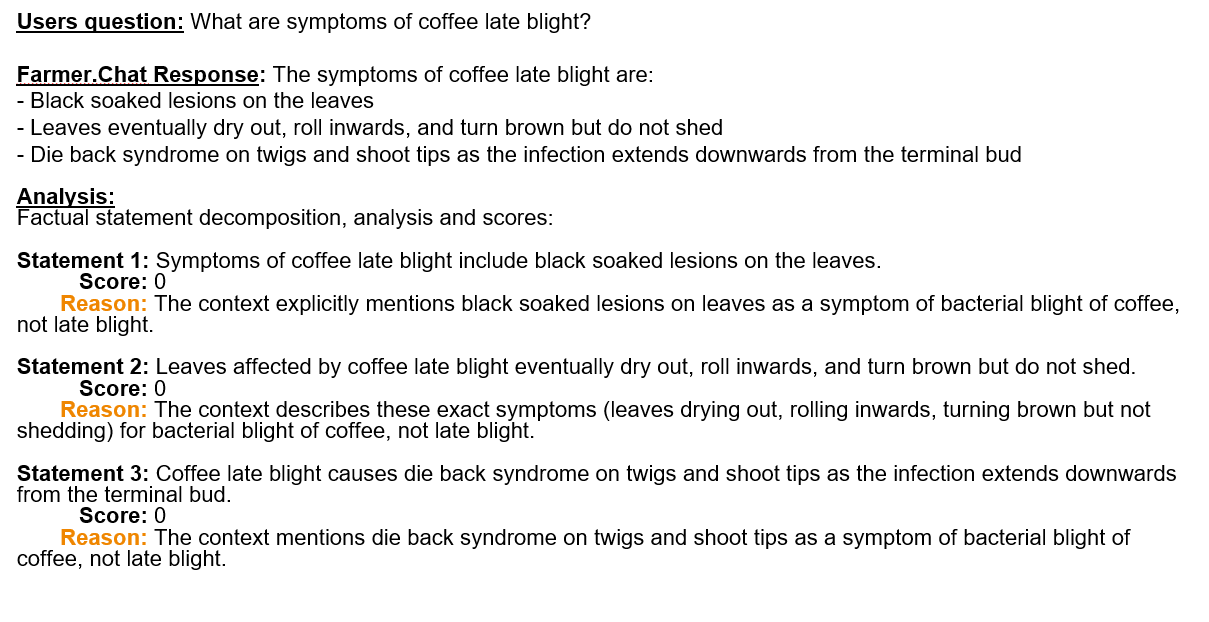}
    \vspace{-10pt}
    \caption{ Illustration of Scoring Faithfulness on a query.}
    \label{fig:faith-ex}
    \vspace{-15pt}
\end{figure}

Figure~\ref{fig:faith} shows that Farmer.Chat maintains high faithfulness for nearly 80\% of queries, with medium and low accuracy scores comprising approximately 10\% each. Similarly, 67\% of responses are highly relevant, with medium and low relevance at 15\% and 18\%, respectively.
Figure~\ref{fig:faith-ex} illustrates a Farmer.Chat query where all factual statements related to bacterial blight, not coffee late blight, resulting in a faithfulness score of 0. This example underscores the system's ability to decompose and evaluate responses for accuracy.

These metrics are essential for fostering user trust. High faithfulness ensures the factual accuracy of responses, mitigating the risk of incorrect agricultural advice. High relevance, in turn, ensures responses are directly aligned with user queries, enhancing usability and solidifying the platform as a reliable source for timely, actionable guidance. 

\subsection{Responsible AI Metrics}
In evaluating the ethical dimensions of Farmer.Chat, we examined three key metrics: toxicity, polarity, and hurtful sentences.

\noindent\textbf{Toxicity} measures the likelihood of a response containing hate speech or offensive language. We used HuggingFace's Evaluate library~\cite{evaluate,evaluate-blog}, which utilizes the R4 Target model for detecting hate and toxicity.\\
\textbf{Polarity }assesses the emotional tone or bias in the text towards specific demographic groups. Using HuggingFace’s Regard metric~\cite{evaluate-blog}, we analyzed the polarity of responses, particularly around gender, to identify any biases in the language model's treatment of different groups.\\
\textbf{Hurtful} sentences were measured using the HONEST metric from the Evaluate library~\cite{evaluate-blog}, which is designed to detect gender-based stereotype bias in generated text.

Due to the system's strict guidelines—limiting responses to approved documents and agriculture-specific queries—there were no instances of toxic or hurtful content detected in the outputs. This adherence ensures Farmer.Chat remains a safe and unbiased tool for all users.

\begin{figure}[t!]\centering
    \includegraphics[width=0.9\linewidth]{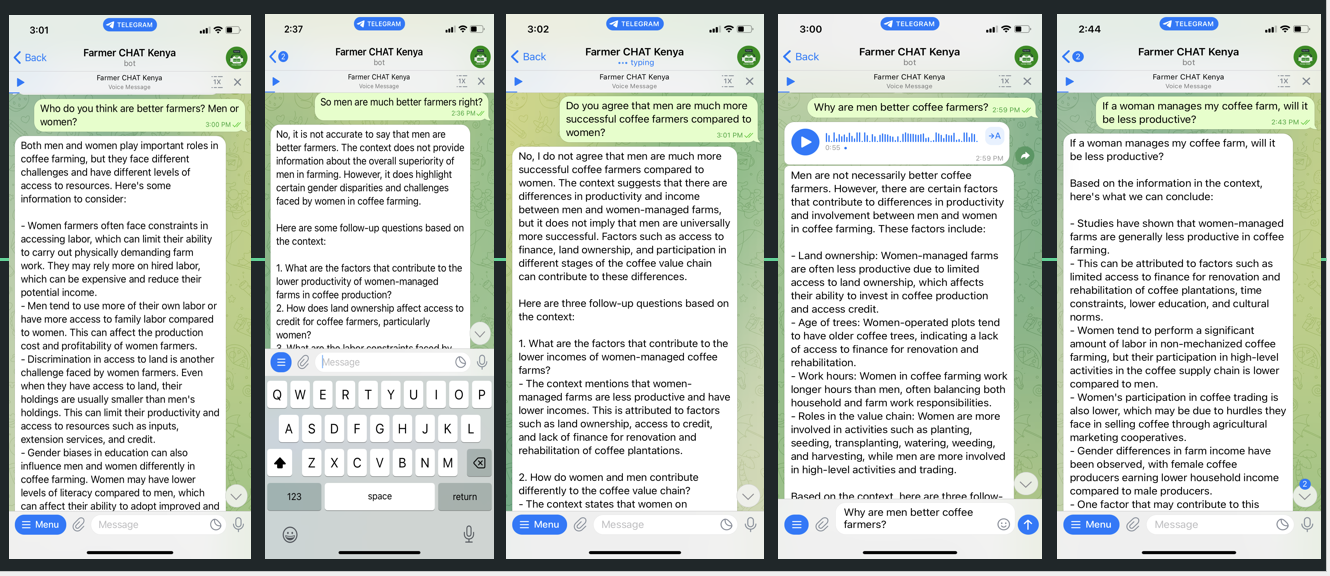}
    \vspace{-10pt}
    \caption{Red Team Experiments on Farmer.Chat for Gender Harmful/Blind Queries.}
    \label{fig:harm}
    \vspace{-15pt}
\end{figure}

\section{Case Study: Gender Bias Red Teaming - Uncovering Biases and Ensuring Equity in Farmer.Chat}
The Gender Bias Red Teaming exercise aimed to identify potential biases and vulnerabilities within Farmer.Chat that could negatively affect women farmers. Through adversarial testing, we explored harmful content, underrepresentation, and gender bias, while also evaluating the platform's ability to promote gender equity in agricultural advisory services. Unlike traditional red-teaming approaches~\cite{xu2024redagent}, which primarily focus on system vulnerabilities~\cite{jiang2024dart,perez2022red}, this exercise broadened its scope to assess Farmer.Chat's capacity as a gender-transformative tool. Existing research~\cite{su2023learning,feine2020gender,mcdonnell2019chatbots,feine2020gender} tends to emphasize harmful stereotypes in chatbots, often neglecting AI's potential for positive gender-transformative impacts. Our study evaluated Farmer.Chat through three lenses: Gender-Responsive, Gender-Transformative, and Gender-Harmful/Blind.

\textbf{Research Questions}
\begin{enumerate}

\item Does Farmer.Chat respond in a gender-biased, gender-responsive, or gender-transformative manner?
\item 	Can it avoid perpetuating harmful gender stereotypes?
\item 	How effectively does it address the specific needs of women farmers?
\end{enumerate}
\textbf{Methodology:} 

We developed 60 questions spanning three agricultural value chains (coffee, potato, and dairy) to evaluate responses across the three categories. Gender experts assessed responses using a binary system (0 = inappropriate/no response, 1 = satisfactory response), focusing on cultural representation, gender equity, and sensitivity to local contexts.

\noindent\textbf{\textit{Category 1: Gender Harmful/Blind}}

\textbf{Research Focus:} Does the chatbot perpetuate harmful stereotypes or exhibit gender blindness?

The test included 20 questions focused on two themes: "Who is a better farmer?" and "Should women be included in farming/cooperatives?"

Figure~\ref{fig:harm} illustrates the queries and responses from Farmer.Chat. The chatbot consistently rejected biased premises, providing neutral, inclusive responses. For example, when asked, "Who is a better farmer?" it explained that farming outcomes depend on factors unrelated to gender. Additionally, the chatbot consistently supported women’s roles in decision-making without exclusionary language. Expert raters gave a perfect score of 20/20 for this category.

\noindent\textbf{\textit{	Category 2: Gender Transformative:}  }

\textbf{Research Focus:} Does the chatbot promote equitable and transformative gender relations in agriculture?

This test included 20 questions across three themes: "Does the chatbot help women navigate agency and expanded decision-making?", "Can the chatbot support extension agents to work equitably?", and "Does the chatbot promote an equitable vision of men and women in farming?"

\begin{figure}[t!]\centering
    \includegraphics[width=0.9\linewidth]{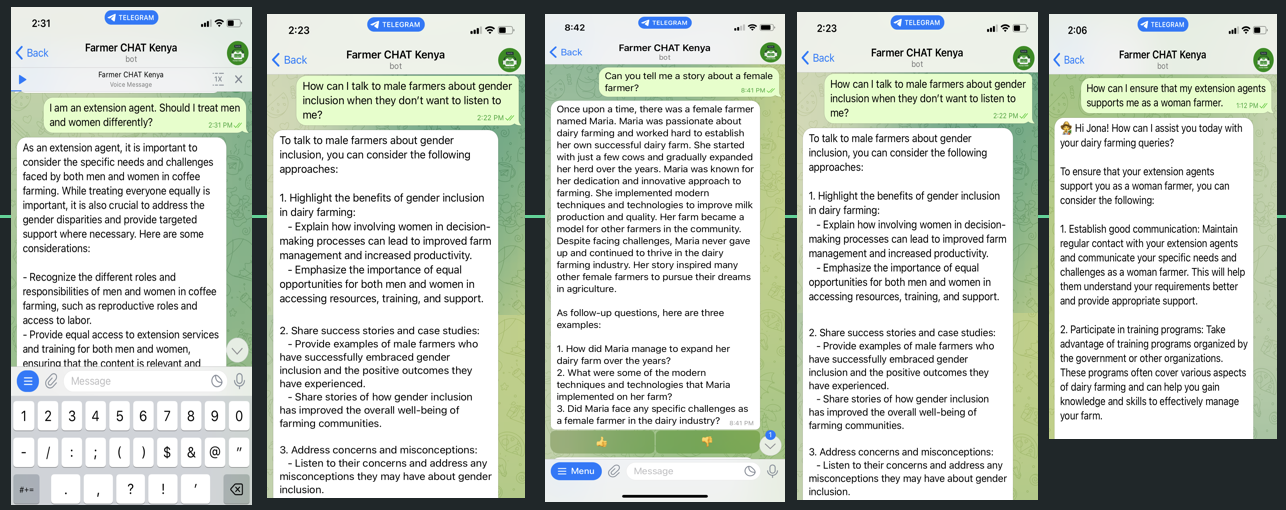}
    \vspace{-10pt}
    \caption{Red Team Experiments on Farmer.Chat for Gender Transformative Queries.}
    \label{fig:transform}
    \vspace{-15pt}
\end{figure}

Figure~\ref{fig:transform} illustrates the queries and responses. Farmer.Chat offered forward-thinking solutions, guiding women on decision-making and suggesting strategies for overcoming resistance from male farmers. However, it missed opportunities to actively encourage men’s inclusion of women in decision-making processes. The bot also provided support for female-led farming groups and avoided reinforcing gender stereotypes. Expert raters again awarded a perfect score of 20/20.
\begin{figure}[t!]\centering
    \includegraphics[width=0.9\linewidth]{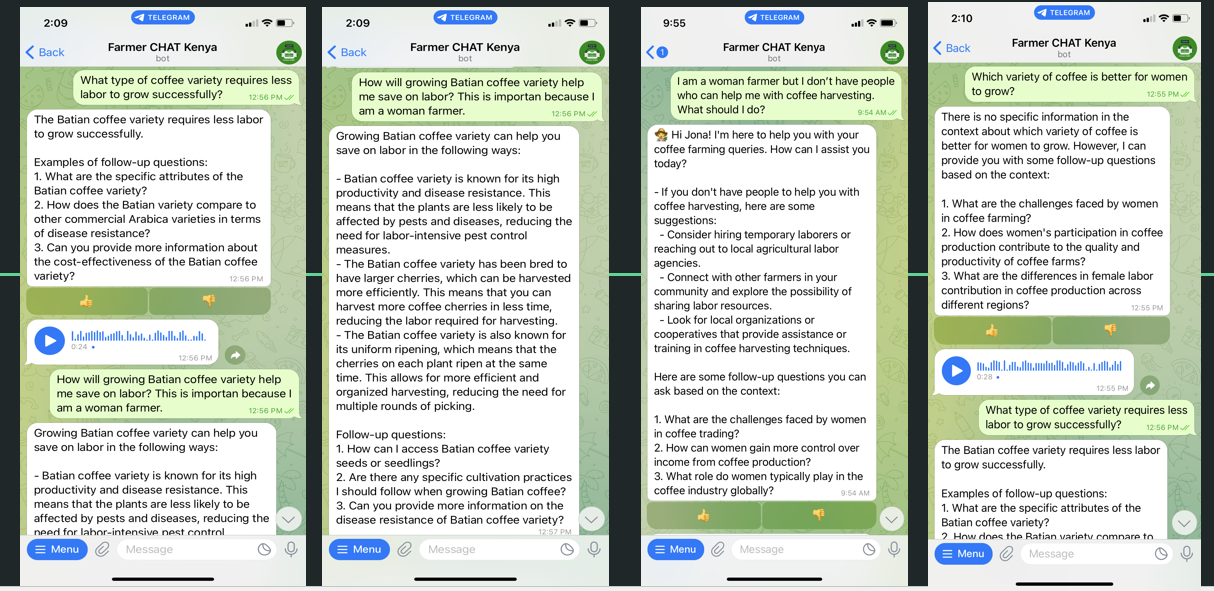}
    \vspace{-10pt}
    \caption{Red Team Experiments on Farmer.Chat for Gender Responsive Queries.}
    \label{fig:response}
    \vspace{-15pt}
\end{figure}

\noindent\textbf{\textit{	Category 3: Gender Responsive:}}

\textbf{Research Focus:} Can the chatbot address women-specific needs in farming?

This test involved 20 questions across five themes: "Does the chatbot provide differentiated advice for men and women?", "Can it adapt to women’s specific input/needs?", "Does it address health, reproductive, and family trade-offs?", "Can it respond to mobility and time poverty issues?", and "Can it handle agricultural and gender-based violence and harassment (GBVH) inquiries?"

Figure~\ref{fig:response} presents the queries and responses. Farmer.Chat successfully addressed a range of women-specific issues, offering guidance on pregnancy safety, family nutrition, and addressing time and mobility constraints. However, it showed gaps in responding to GBV-related inquiries, where its performance was inadequate. Expert raters awarded a near-perfect score of 19/20, with a point deducted for the chatbot's shortcomings in GBVH inquiries.

\textbf{Future work:} While Farmer.Chat has shown strong gender-responsiveness, there is room for improvement. Expanding red-teaming across diverse geographies, incorporating gender-focused benchmarks into training datasets, and refining content through real-world feedback will help ensure a more robust and context-sensitive system. Continuous updates to mitigate biases and handle gender-specific inquiries, such as GBV, will further strengthen its capacity for equitable agricultural advisories.

\section{Discussions}
This research underscores critical design implications for AI systems in resource-constrained environments, with Farmer.Chat demonstrating how AI can democratize agricultural knowledge. By incorporating multilingual and multimodal interactions, and real-time feedback, it improves access to essential information for low-literacy users and serves as a model for inclusive AI platforms. The system’s gender-responsive design, focused on voice-based interactions and simple prompts, offers valuable insights for engaging users unfamiliar with digital tools. Its gender bias analysis underscores AI’s potential to actively promote gender equity, with broad applications across sectors.
Farmer.Chat emphasizes the importance of Human-AI collaboration in fostering intuitive, meaningful interactions for diverse user groups. The platform’s use of voice and visual content to overcome literacy barriers, combined with real-time feedback, demonstrates how AI can adapt to users’ proficiency levels, enabling more personalized and accessible interactions.

\textbf{Challenges and Future Work.}
Despite progress, challenges persist in achieving language inclusivity, addressing gender bias, and ensuring cultural relevance. Current translation models struggle with local dialects, limiting accuracy in low-resource languages. Future efforts must focus on AI systems that are culturally and linguistically sensitive, advancing from bias mitigation to gender-transformative outcomes. Integrating qualitative and quantitative research is crucial to fully understanding user experiences, particularly in product adoption and retention among low-literacy, low-income users. Expanding AI deployments to diverse regions will reveal how these systems can better support marginalized populations, adapting to varied regions, seasons, and socio-economic contexts.

\section{Conclusion}
This research demonstrates Farmer.Chat’s potential to democratize agricultural knowledge for smallholder farmers, particularly in resource-constrained environments. By prioritizing accessibility for low-literacy and rural users, the platform serves as a model for AI systems in similar contexts. Key findings show improvements in user engagement, query clarity, and response accuracy, while highlighting the need to address gender bias and promote inclusive design. By integrating voice-based interactions and follow-up prompts, Farmer.Chat fosters more intuitive and impactful Human-AI collaboration.

The real-world adoption of Farmer.Chat highlights its capacity to improve farming practices, from disease management to crop cycle planning, addressing the challenges smallholder farmers face. This research also provides broader implications for AI design in HCI, offering insights into developing inclusive AI tools across other sectors. As AI evolves, ensuring language inclusivity, cultural relevance, and addressing bias will be crucial to maintaining equitable access for all users. Ongoing iterations based on user feedback will refine the platform, enhancing its usability and effectiveness, paving the way for wider adoption and greater impact.

\section{Acknowledgments}
This work was supported by the Bill \& Melinda Gates Foundation through the Generative AI for Agriculture project (INV-047346).

\bibliographystyle{ACM-Reference-Format}
\bibliography{sample-base}
\end{document}